\documentclass{ws-p9-75x6-50}

\begin{document}

\title{Geodetic Precession in Binary Neutron Stars}

\author{Michael Kramer}

\address{University of Manchester, Jodrell Bank Observatory, \\
Macclesfield, Cheshire SK11 9DL, UK 
\\ E-mail: mkramer@jb.man.ac.uk}


\maketitle

\abstracts{ The most recent evidence for geodetic precession observed
in binary radio pulsars is presented and discussed. It is demonstrated
how an analysis of these results can be used to study theories of
gravity, stellar evolution and pulsar emission theory.  In order to
highlight the observational strategies, an overview of the applied
techniques is given.  }
\section{Introduction}

The discovery of pulsars in 1967\cite{hbp+68} marked the beginning of
a new era in astrophysics. Besides studying pulsars as fascinating
astrophysical objects in their own right, these sources serve as
invaluable tools in the research of a large variety of physical and
astrophysical problems.  Studied subjects range from plasma and
solid-state physics under extreme conditions to investigations of the
interstellar medium. One can study stellar and binary evolution as
well as high precision astrometry and planetary ephemerides.
Applications are found in cosmology, quantum physics and, of course,
in gravitational physics.  The latter application became possible with
the discovery of the first binary pulsar by Hulse \& Taylor
in 1974\cite{ht75a}.  As they immediately realized, a
pulsar in a binary orbit represents a highly stable and accurate clock
orbiting in the gravitational field of (usually) compact stars. They
were fortunate to discover in PSR B1913+16 the first double neutron
star system. These systems are doubly rare because they are formed
less often and also since they are more difficult to detect
(significant Doppler smearing of the pulse signal occurs when the
orbits are less than about 1 day).

Meanwhile many more binary pulsars have been discovered. At the time of
writing about 5\%, or 70 pulsars out of more than 1300 known, are in binary
systems. 
Three pulsars are in orbit with normal or giant stars, and
two pulsars are members of planetary systems. The rest of the binary pulsar
population has compact stars, i.e.~white dwarfs or neutron stars, as 
companions. The majority are orbited by white dwarfs of
typically $0.2M_\odot$, although there is a small but growing population of
heavy white-dwarf companions\cite{klm+00a,eb00,clm+01}. 
Pulsar--white-dwarf systems
have been used to perform unique tests of certain aspects of gravitational
theories\cite{BD96}, for instance testing the existence of gravitational
dipole radiation as predicted by many alternative theories of
gravity\cite{lcw+01} or the violation of the strong equivalence
principle\cite{wex00}. However, for tests of theories of gravity in a strong
field limit, observations of double neutron star systems still yield the most
exciting results.

Only five double neutron star systems (DNSs) are known. Orbital periods range
from 7.7 hours for PSR B1913+16\cite{ht75a} to 18.8 days in the case of the
most recent discovery, PSR J1811$-$1736\cite{lcm+00}. Presently, two of the
five DNSs allow the determination of more than two post-keplerian (PK)
parameters. The PK parameters describe relativistic corrections to 
Newtonian physics, represented by the five standard Keplerian parameters, and
can be written as functions of the pulsar and companion masses, $m_p$ and
$m_c$, and the Keplerian parameters. These functions will differ for different
theories of gravity, but when Keplerian and PK parameters are {\em measured by
observations in a manner independent of a particular theory of gravity}, the
observations can be compared to theory\cite{dt92}.  If two PK parameters are
measured, a given theory of gravity will produce values for the pulsar and
companion mass.  More generally, a measurement of $n$ PK parameters describes
$n$ curves in a two dimensional $m_p$-$m_c$ plane whose shape and position
depend on the theory of gravity being applied.  If the chosen theory is an
accurate and adequate description of the physics, all $n$ curves should meet
in a single point. A measurement of at least three PK parameters hence allows
one to perform tests of gravitational theories

The measurement of PK parameters has been possible thanks to a
powerful technique
called {\em pulsar timing}. When giant radio telescopes
are used to measure the arrival time of photons emitted by the pulsar,
a breath-taking accuracy can be achieved in experiments which are
impossible to perform on Earth or in the solar system. For details, we
refer to the contribution by Stairs (these proceedings) and to
reviews like that of Backer \& Hellings (1986)\cite{bh86}.  
Despite the remarkable
successes achieved with pulsar timing, one particular aspect of
gravitational physics can only be studied adequately if the 
structure of the received pulses itself is analysed carefully.

In general relativity, the proper reference frame of a freely falling object
suffers a precession with respect to a distant observer, called geodetic
precession. A direct measurement of this effect is the scientific goal of the
Gravity Probe-B satellite experiment (see Everitt, these proceedings).  In a
binary pulsar system this geodetic precession leads to a relativistic
spin-orbit coupling, analogous to spin-orbit coupling in atomic physics.  As a
consequence, the pulsar spin precesses about the total angular momentum,
changing the relative orientation of the pulsar towards Earth. In such a case,
we should expect a change in the radio emission received from the
pulsar, as first proposed by Damour \& Ruffini\cite{dr74} very soon after the
discovery of PSR B1913+16. Hence, in contrast to ``only'' accurately
registering the arrival times of pulsar signals, we are even more interested
in the emission properties when we want to study geodetic precession.

In this paper we will summarize how to study geodetic precession in binary
pulsars. As we have to analyse the emission properties of radio pulsars, we
first provide a brief introduction into pulsars and their radiation
properties. We report observations for PSR B1913+16 and discuss models 
used to
describe detected pulse shape changes. We demonstrate how information about
the system geometry can be used to learn about the effects of asymmetric
supernova explosions and ``kicks'' imparted to newly born neutron
stars. Observations of two more binary pulsars, PSRs B1534+12 and
J1141$-$6545 are discussed as they provide more evidence for geodetic
precession.

\begin{table}[t]
\caption{Selected parameters for the pulsars discussed. Uncertainties
in the last digits are quoted in brackets.\label{tab:parms}}
\begin{center}
\footnotesize
\begin{tabular}{|lrrr|}
\hline
& \multicolumn{1}{c}{PSR B1913+16$^1$} & \multicolumn{1}{c}{PSR B1534+12$^2$} & 
  \multicolumn{1}{c|}{PSR J1141$-$6545$^3$} \\
\hline
$P$ (ms) & 59.029997929613(7) & 37.9044403237164332(10) & 393.897833900(3)\\
$P_b$ (d) & 0.322997462736(7) & 0.4207373013796296135(10) &  0.197650966(6) \\
$e$ & 0.6171308(4) & 0.2736777(5) & 0.171881(9) \\
$a\;\sin i$ (lt-s) & 2.3417592(19) & 3.729463(3) & 1.859470(14) \\
$m_p$ (M$_\odot$) & 1.4411(7)  & 1.339(3) & \multicolumn{1}{c|}{?} \\
$m_c$ (M$_\odot$) & 1.3879(7)  & 1.339(3) & \multicolumn{1}{c|}{?} \\
\hline
\end{tabular}

$^1$Taylor \& Weisberg (1989), 
$^2$Stairs et al.~(1998), $^3$Kaspi et al.~(2000)
\end{center}
\end{table}

\section{Pulsars}

\subsection{The Clock}

Pulsars are rotating neutron stars. From timing measurements of binary radio
pulsars we determine the masses of pulsars to be within a quite
narrow range of
$1.35\pm0.04 M_\odot$\cite{tc99}. Model calculations involving different
equations of state produce results for the size of a neutron star quite
similar to the very first calculations by Oppenheimer \& Volkoff
(1939)\cite{ov39}, i.e.~about 20 km in diameter.  Such sizes are consistent
with independent estimates derived from modelling light-curves and
luminosities of pulsars observed in X-rays, e.g.~for PSR
J0437$-$4715\cite{pz97}.

These massive and compact stars rotate with typical periods of about 0.6 s
for ``normal'' pulsars, and periods as low as 1.56 ms for millisecond
(``recycled'') pulsars (cf.~Fig.~\ref{fig:ppdot}). 
Considering that this smallest rotational period for
PSR B1937+21 corresponds to more than 38000(!)  revolutions per minute makes
the large stability of pulsars as clocks easy to understand.

Typical magnetic field strengths at neutron star surfaces are
of order $10^{12}$ G. By comparison, the Earth's magnetic field is
only 1 G. We can infer these field strengths from the slow increase
in pulse period due to the loss of rotational energy by the emission
of magnetic dipole radiation, or more directly from synchrotron lines
in X-ray spectra of X-ray pulsars\cite{bu78}.

\begin{figure}[t]

\begin{minipage}{6cm}
\epsfxsize=17pc 
\epsfbox{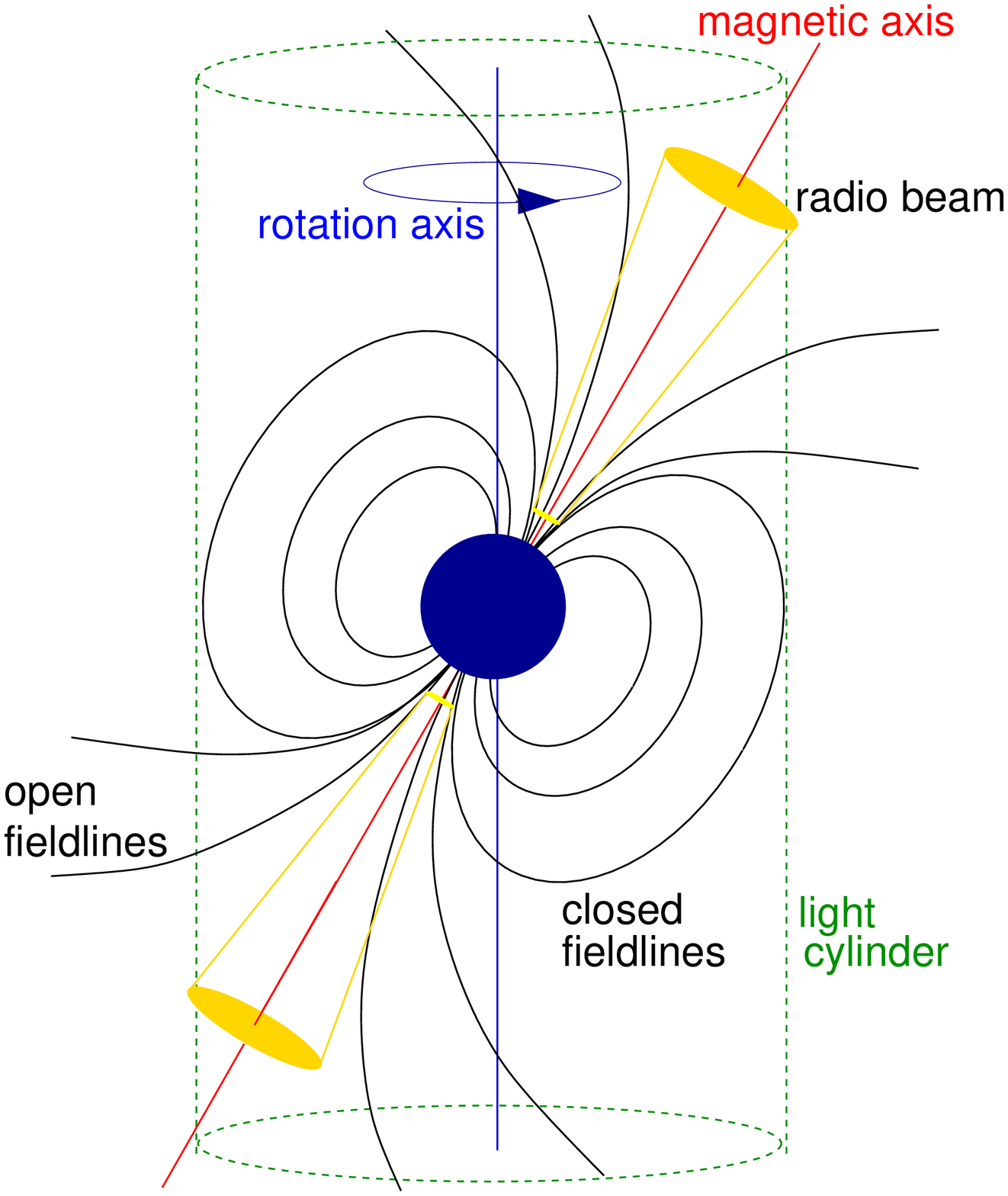} 
\end{minipage}
\hfill
\begin{minipage}{6cm}
\epsfxsize=16pc 
\epsfbox{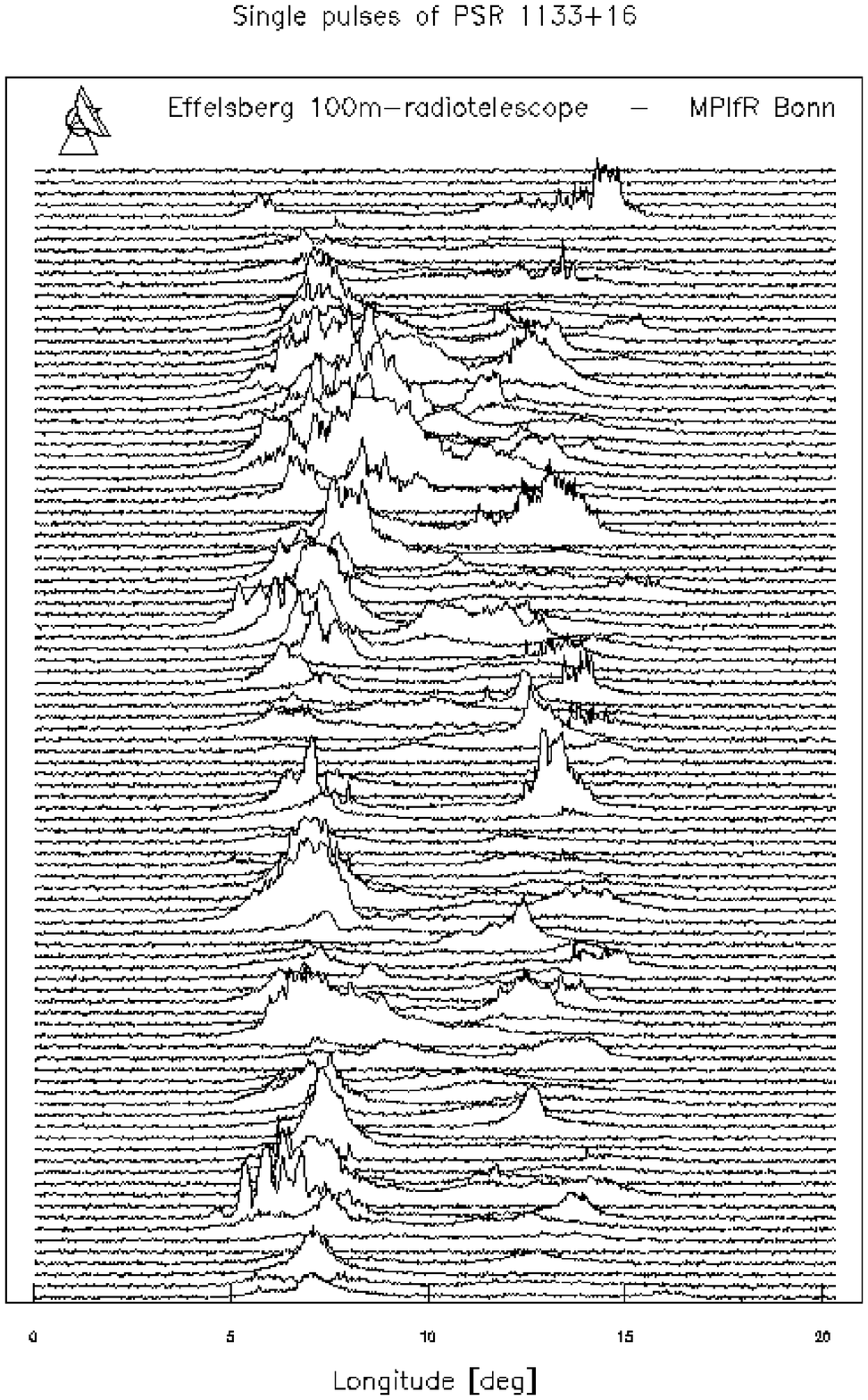} 

\epsfxsize=13.5pc 
\hspace{0.5cm}\epsfbox{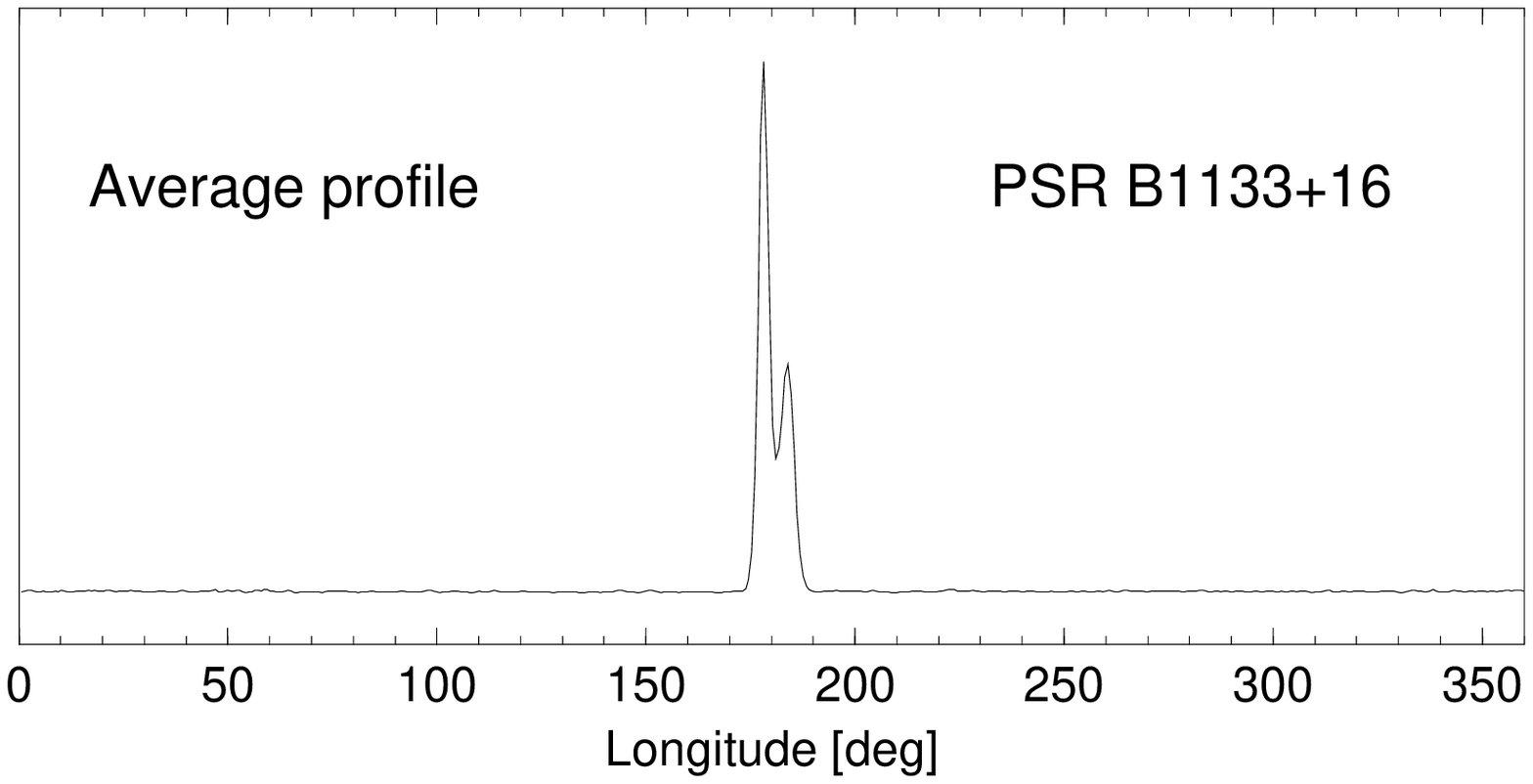} 
\end{minipage}

\caption{left) A pulsar is a rotating, 
highly magnetised neutron star. A radio beam
centred on the magnetic axis is created at some distance to the pulsar.
The tilt between the rotation and magnetic axes makes the pulsar 
in effect a cosmic lighthouse when the beam sweeps around in space;
right) while individual pulses vary in shapes and strength (zoomed in,top),
average profiles are stable (full profile, bottom). 
The typical pulse duty cycle 
is only $\sim$4\%.
\label{fig:1133}\label{fig:pulsar}}
\end{figure}

\subsection{The Clockwork}

The tick of the pulsar clock is provided by a narrow radio beam centred
on the magnetic axis of the pulsar, which is inclined to the rotation
axis (see Fig.~\ref{fig:pulsar}). If the pulsar beam is directed
towards Earth once per rotation, radio telescopes on Earth may be able
to pick up a pulsed radio signal. This periodic beacon sent by the
pulsar clock is usually rather weak due to the distance of the pulsar,
but also due to the small size of the emission region, which we expect
to be only a few hundred km across\cite{kxj+96}.  Nevertheless, the
radio emission is actually rather intense as we measure brightness
temperatures exceeding $10^{30}$ K, which we can only explain by
assuming a coherent emission mechanism. After 35 years of pulsar
research, the details of the 
actual emission process still elude us, but at least we
have some basic understanding sufficient to perform the experiments
described later.

As the neutron star rotates with its magnetic field, an induced
electric quadrupole field pulls out easily charges from the stellar surface
(the electrical force exceeds the gravitational force by 
a factor of $10^{12}$!),
surrounding the pulsar with dense plasma.  The magnetic field forces the
plasma to co-rotate with the pulsar like a rigid body. This co-rotating {\sl
magnetosphere} can only extend up to a distance where the
co-rotation velocity reaches the speed of light\footnote{Strictly speaking,
the Alfv\'en velocity will determine the co-rotational properties of the
magnetosphere.}. This distance defines the so-called light cylinder which
separates the magnetic field lines into two distinct groups, i.e.~{\sl open
  and closed field lines}. Closed field lines are those which close within the
light cylinder, while open field lines would close outside. The plasma on the
closed field lines is trapped and will co-rotate with the pulsar forever. In
contrast, plasma on the open field lines can 
reach highly relativistic velocities and can 
leave the magnetosphere, creating
the observed radio beam at a distance of a few tens to hundreds of km above
the pulsar surface (see Fig.~\ref{fig:pulsar}).

\subsection{The ticks}

\smallskip

\noindent
{\it Individual Pulses and Average Pulse Profiles}

\medskip

\noindent
Individual pulses reflect the instantaneous plasma processes in the
pulsar magnetosphere at the moment when the beam is directed towards
Earth. The dynamics of these processes results in often seemingly
random individual pulses, in particular when viewed with high time
resolution (see Fig.~\ref{fig:1133}).  Despite this variety displayed
by the single pulses, the mean pulse shape computed by averaging a few
hundreds to few thousands of pulses is incredibly stable\cite{bla91}.
In contrast to the snapshot provided by the individual pulses, the
average pulse shape, or {\em pulse profile}, can be considered as a
long-exposure picture, revealing the global circumstances in the
magnetosphere. These are mostly determined by geometrical factors and
the strong magnetic field, leading to very stable pulse profiles.
Apart from a distinct evolution with radio frequency, the same
profiles are obtained, no matter where and when the pulses used to
compute the average have been observed.  \footnote{ For completeness
we should note that there is a very small fraction of
pulsars\cite{bac70,lyn71a,kxc+99} exhibiting profile changes on short
time scales of minutes or hours.  In these cases, the profiles do not
exhibit random shapes either, but the pulsar seems to switch between a
few, usually two, distinct profiles. The origins of these phenomena
are not understood. On secular time scales, these pulse profiles are
still stable.}

\bigskip

\noindent
{\it Beam Shapes}

\smallskip 

\begin{figure}[t]


\begin{tabular}{cc}
\epsfxsize=13pc 
\epsfbox{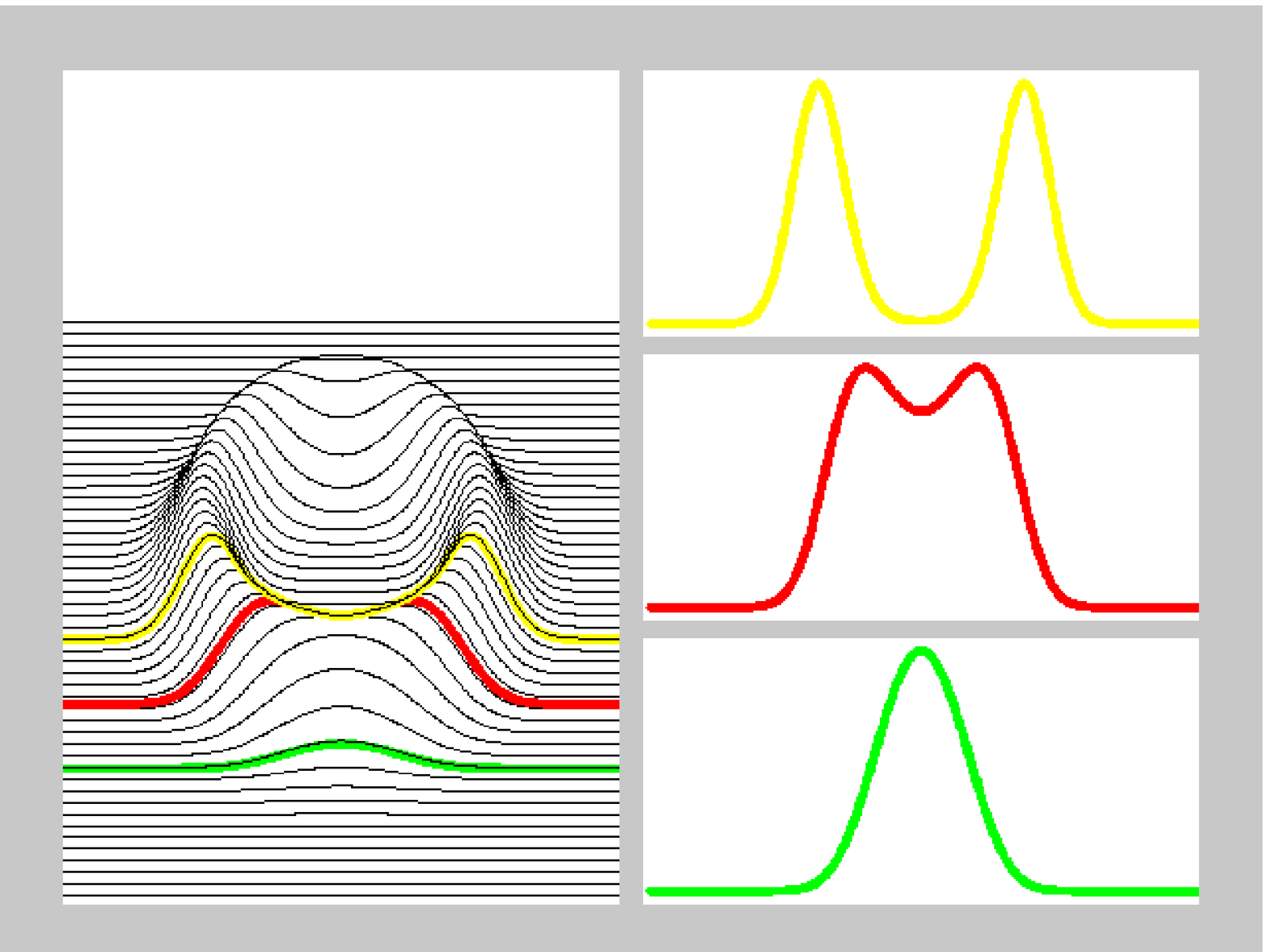} & 
\epsfxsize=13pc 
\epsfbox{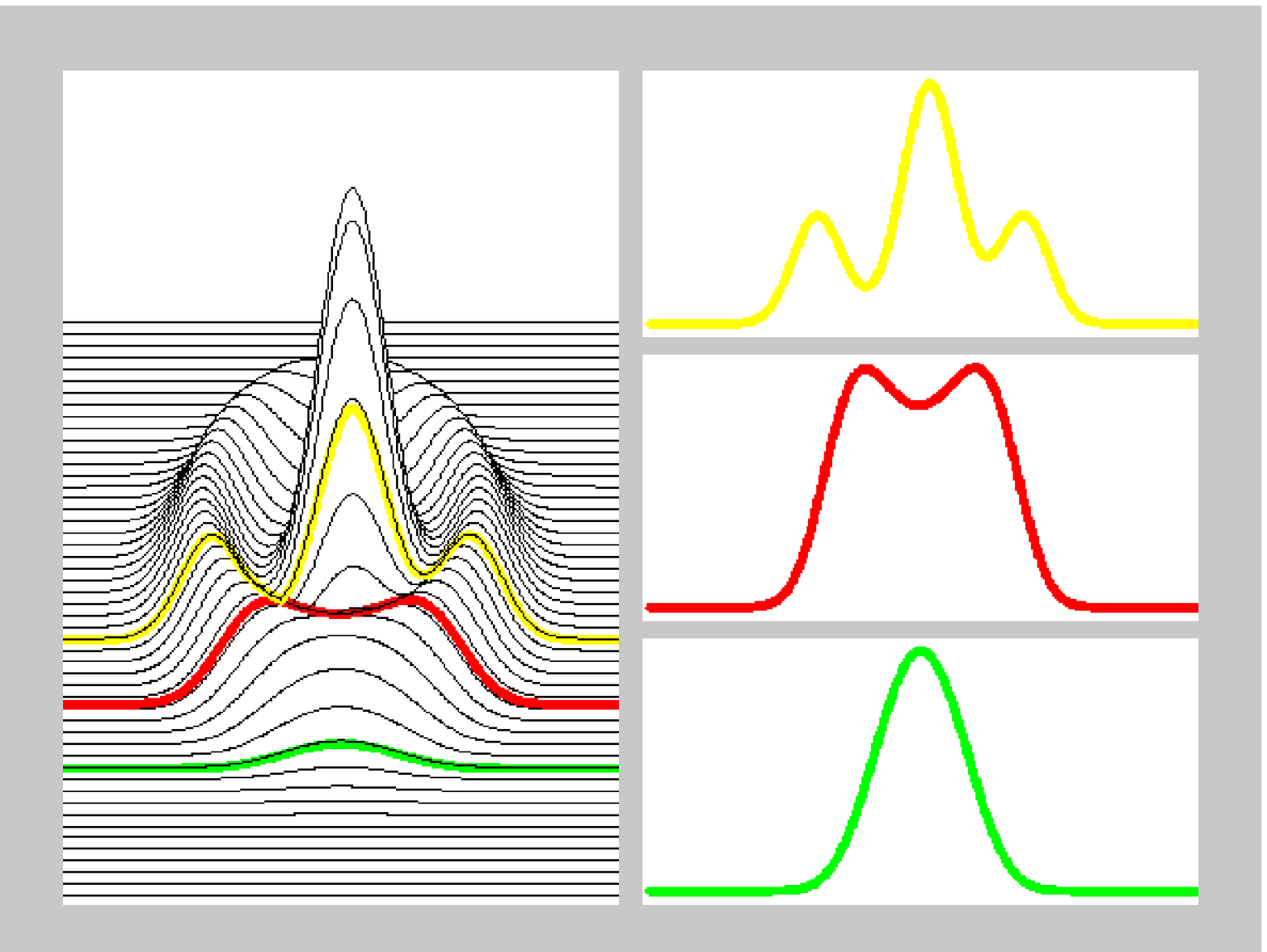} 
\end{tabular}

\caption{left) Hollow cone model to explain the observed
pulse profiles as different cuts through the emission beam;
right) very often an additional ``core'' component is observed near the
centre of the beam, in particular at low frequencies.
\label{fig:shape}}
\end{figure}

\noindent
The observed pulse profiles show a large variety of shapes. Although each
pulsar exhibits a slightly different profile -- almost like a unique
fingerprint -- a systematic pattern can be recognized. The most simple model
successfully describing the beam shapes is that of a hollow cone of 
emission\cite{kom70}. It is based on the idea that the outermost open
field lines, which show the largest curvature among the ``emitting field
lines'', should be associated with the strongest emission, leading naturally
to a cone-like structure. Observations show that this picture is
oversimplified, since we also often observe components near the centre of
the cone, a so called {\em core} component, in particular at lower
frequencies.  However, the combination of a hollow cone with a core component
can be successfully applied to understand the vast majority
of pulsar profiles\cite{os76b,ran83,kwj+94}.  Depending on the way how
our line-of-sight cuts the emission cone, different pulse profiles are
observed (Fig.~\ref{fig:shape}).

\bigskip

\noindent
{\it Polarisation --- Signatures of Geometry}

\smallskip 

\noindent
An important and most useful property of pulsar radio emission is its typical
high degree of polarisation. The radiation is often 100\% elliptically
polarized, and it is usual practice to separate the polarisation into 
linearly and circularly polarized components. The linear component is 
often the
far dominating one, although pulsars with circular components as strong as
30\% or more are not uncommon. 
The degree of polarisation decreases with increasing
frequency, leaving pulsars more or less completely de-polarized at high
frequencies\cite{xkj+96}. At low frequencies, however, the polarisation
serves as a useful diagnostic tool, both to measure the magnetic field of the
ionized interstellar medium and, in particular for our purposes, to obtain
information about the viewing geometry.

The strong coupling of the 
outwards-moving plasma to the magnetic field lines in
the pulsar magnetosphere has the effect that the plane of polarisation of the
linear component is determined by the plane embedding the corresponding field
line. The observed position angle (PA) of the linear polarisation is
then given by the projection of this direction onto our line-of-sight. The
result is an S-like curve of the PA whose shape depends on the angle between
the rotation and magnetic axes as well as on the distance of our line-of-sight
to the magnetic pole. If our line-of-sight cuts the emission beam close to the
magnetic axis, the PA changes rapidly when crossing the centre of the
beam. If we are cutting the cone further away from the pole, the transition is
much smoother and the PA swing much flatter.

By measuring the polarisation characteristics of a pulsar, we can in principle
win information about the pulsar's orientation
towards us. In practice, fitting this {\em rotating vector model}\cite{rc69a}
(RVM) often turns out to be difficult. Although the majority of observed PA
swings can be well described by the RVM after correcting for sometimes
occurring orthogonal modes (i.e.~jumps of the PA by nearly 90$^\circ$ which
are probably magnetospheric propagation effects), the
uncertainties in the obtained angles representing the geometry are typically
large. The reason is not a failure of the model, but simply the small 
size of beam of most
pulsars, which provides constraints to the fit for only the small fraction of
the pulse period when the pulse is actually observed, which is typically only
4\%.

\begin{figure}[t]


\begin{tabular}{cc}
\epsfxsize=13pc 
\epsfbox{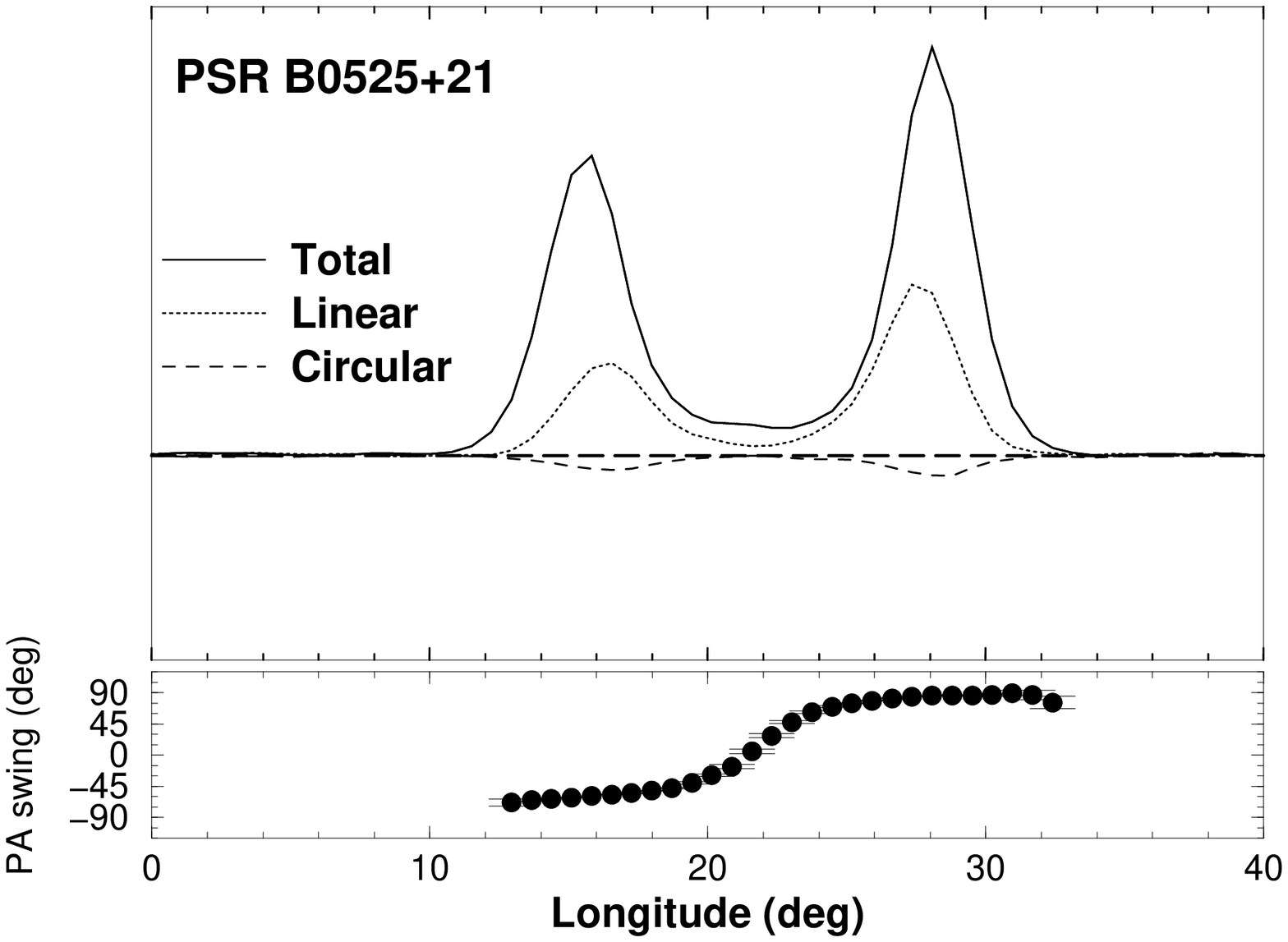} & 
\epsfxsize=13pc 
\epsfbox{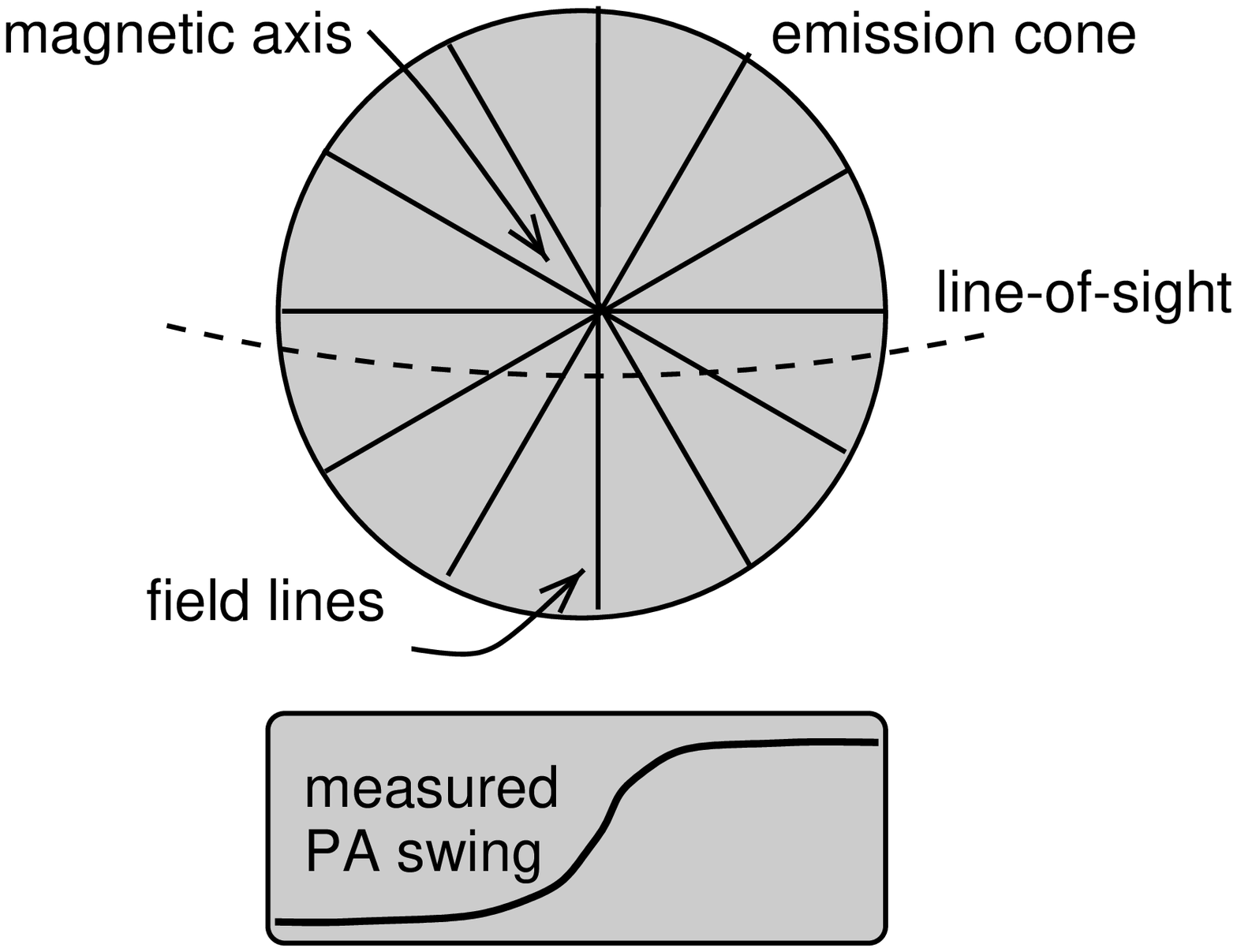} 
\end{tabular}

\caption{left) Pulsar emission is often highly polarized. The PA
of the linear component usually performs an S-like swing across the profile,
right) the change in PA can be understood as a projection of the field
line direction onto our line-of-sight.
\label{fig:rvm}}
\end{figure}

The geometrical origin of the PA swing across the pulse is
impressively demonstrated by its independence upon observing
frequency\footnote{after correcting for propagation effects due to the
interstellar medium, i.e.~Faraday rotation}. Even if the results
of a RVM fit to the PA data are not very well constrained, an {\em
observed change} in the swing curve will immediately indicate a change
in viewing geometry.

\section{Evidence for Geodetic Precession --- Part I.}

Immediately after the discovery of the 59-ms binary pulsar
PSR B1913+16 it was realized that the
system should exhibit a measurable amount of geodetic precession if the pulsar
spin axis is misaligned with respect to the orbital angular momentum vector
\cite{dr74,bo75,bo75b,eh75}
We will discuss the possible origin for a misalignment  of the
spin vectors later and first concentrate on the expected observational
consequences.

\subsection{Precession rate}

The precession rate is given by\cite{bo75b,ber75}
\begin{equation}
\Omega_p  = \left( \frac{2\pi}{P_b}\right)^{5/3} \cdot
  T_\odot^{2/3} \cdot
  \frac{m_c(4m_p+3m_c)}{2(m_p+m_c)^{4/3}} \cdot
  \frac{1}{1-e^2}
\end{equation}
where $P_b$ is the period and $e$ the eccentricity of the orbit. We express
the masses $m_p$ and $m_c$ in units of solar masses ($M_\odot$) and the define
the constant $T_\odot=GM_\odot/c^3=4.925490947 \mu$s. $G$ denotes the
Newtonian constant of gravity and $c$ the speed of light. 

The values for the orbital period and the eccentricity of the PSR
B1913+16 system can be obtained from timing observations (see Table
\ref{tab:parms}).  These observations have also resulted
in the measurements of three PK parameters, providing the 
most stringent
test of gravitational theories in the strong field limit so
far\cite{tw89,dt91,tay94b}. General relativity has passed all these
tests with flying colours. We therefore have safe and accurate
measurements for the masses of both pulsar and its companion:
$m_p=1.4411\pm0.0007$ and $m_c=1.3879\pm0.0007$. (Strictly speaking
these {\em observed} masses differ from the intrinsic ones by an
unmeasurable Doppler factor due a radial velocity, which cannot be
determined from timing measurements. We can safely neglect this small
difference here.) Using these values we obtain a precession
rate of $\Omega_p=1.21$ deg yr$^{-1}$.  Since the orbital angular
momentum is much larger than the pulsar spin, the orbital spin
practically represents a fixed direction in space, defined by the
orbital plane of the binary system. Given the calculated precession
rate, it takes 297.5 years for the pulsar spin vector to precess
around it.

\subsection{First Signs -- First Riddles}

\begin{figure}[t]


\centerline{
\begin{tabular}{cc}
\epsfxsize=12pc 
\epsfbox{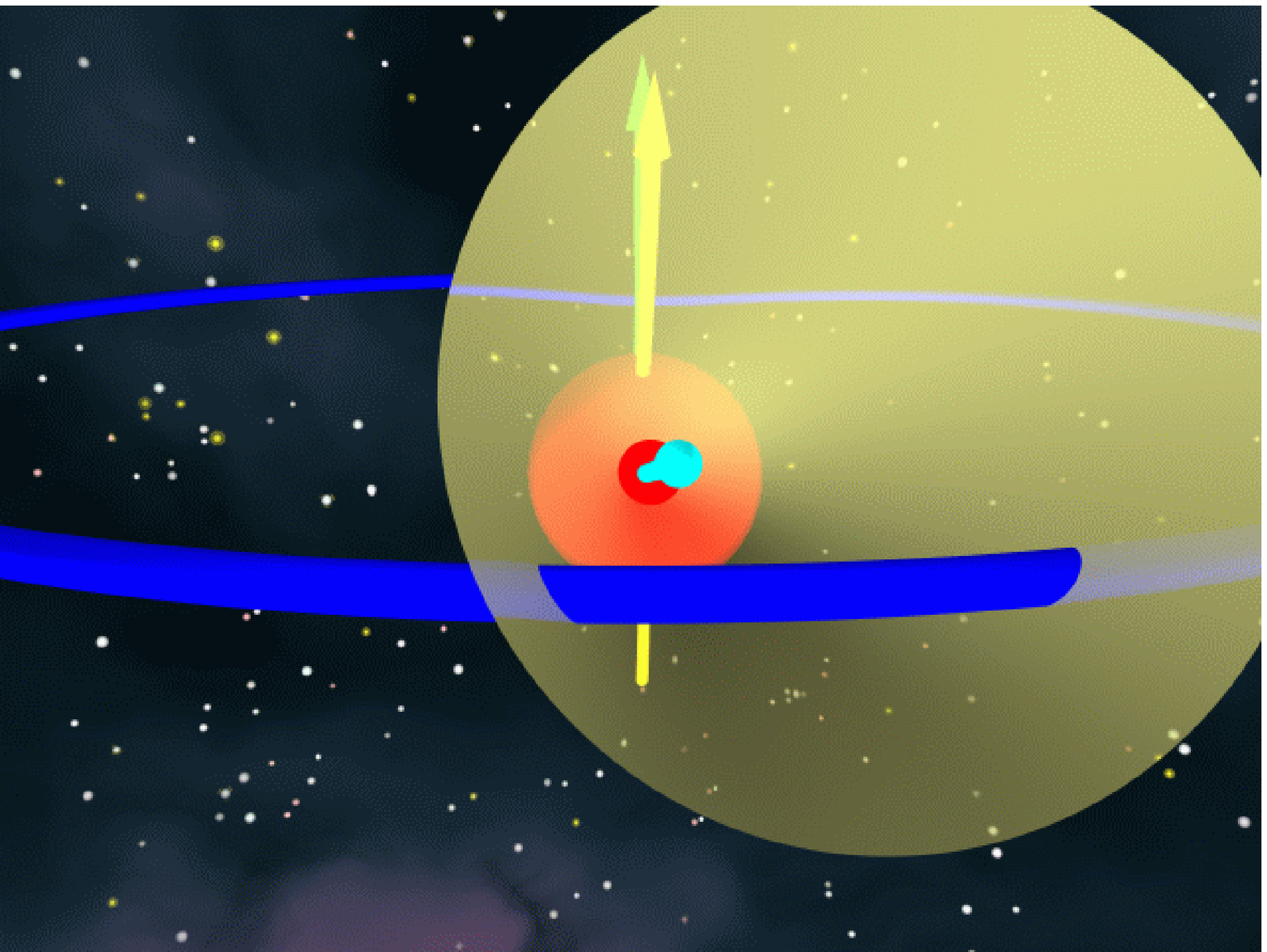} & 
\epsfxsize=12pc 
\epsfbox{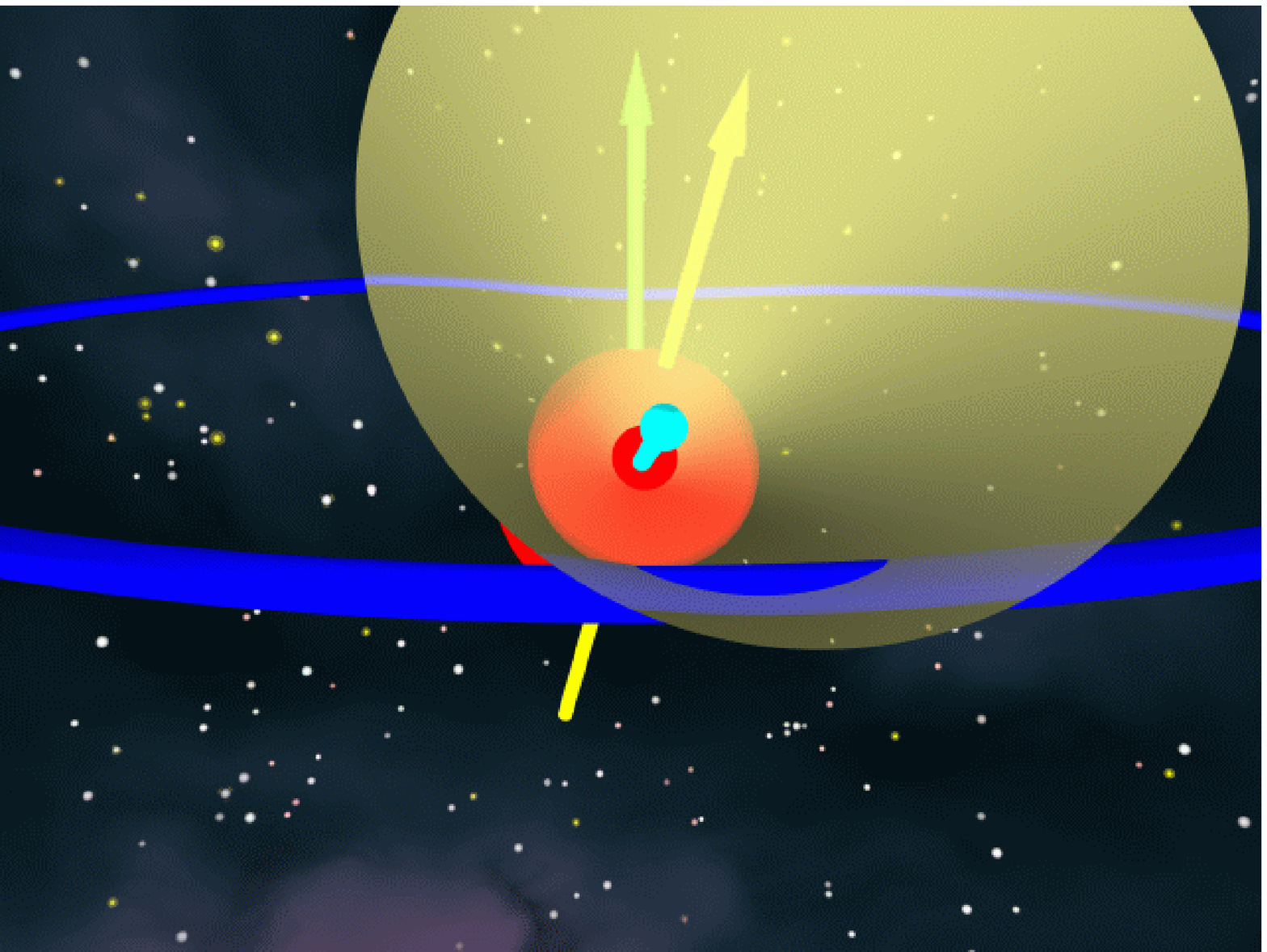} 
\end{tabular}
}

\caption{Precession changes the cut of our line-of-sight (indicated by
circle) through the emission cone. The consequence should be a
change in the width and possibly also shape of the pulse profile.
Since the distance between line-of-sight and magnetic axis is changing,
an alteration of the position angle swing is expected.
\label{fig:slices}}
\end{figure}

As a result of the precession the angle between the pulsar spin axis and our
line-of-sight should change with time, so that different portions of the
emission beam are observed. Consequently, one expects changes in the measured
pulse shape, in particular in the profile width, as a function of time (see
Fig.~\ref{fig:slices}). In the extreme case, the precession may move the beam
out of our line-of-sight and the pulsar may disappear from the sky until it
becomes visible again.

\begin{figure}[t]


\centerline{
\begin{tabular}{cc}
\epsfxsize=12.5pc 
\epsfbox{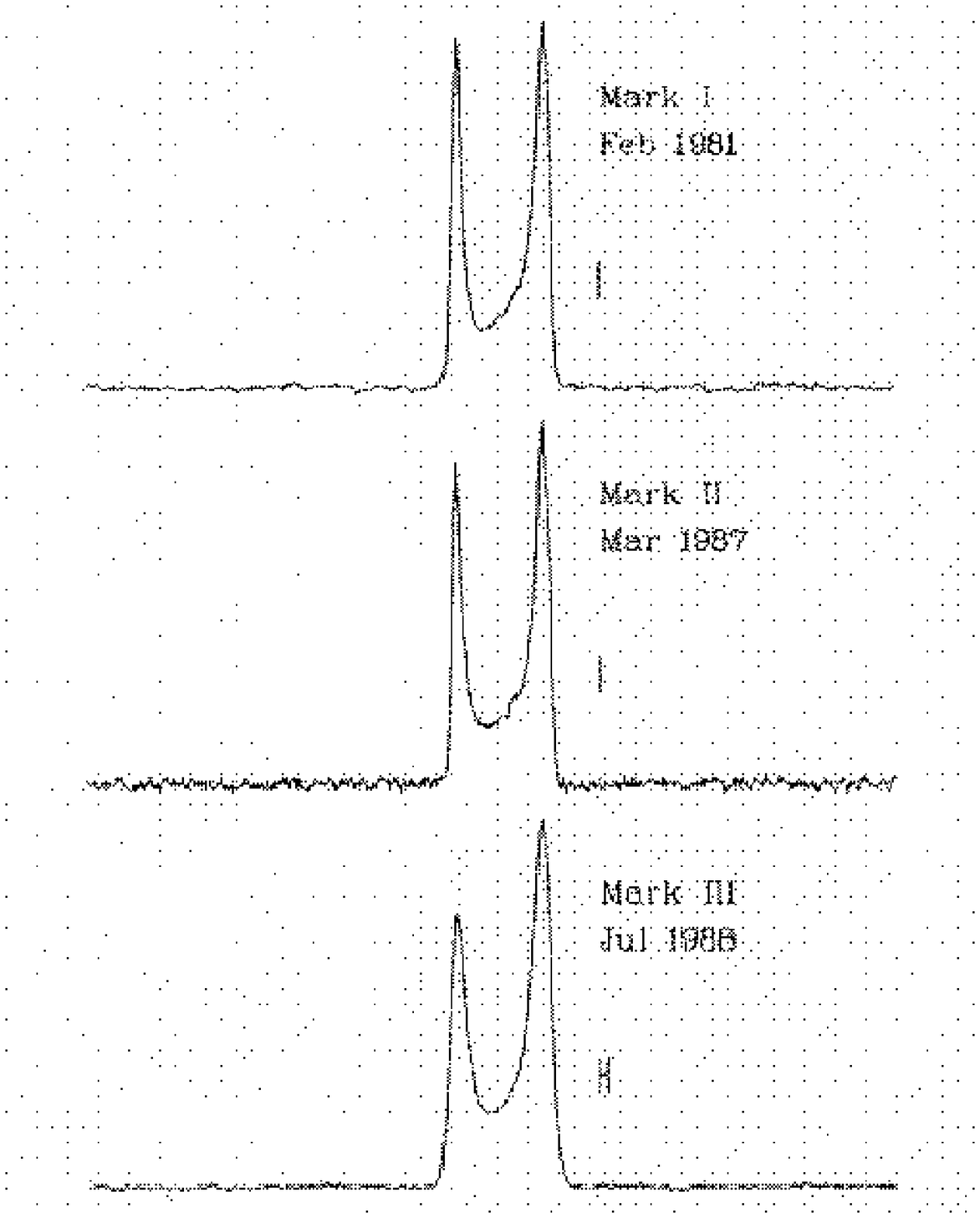} 
\epsfxsize=11pc 
\epsfbox{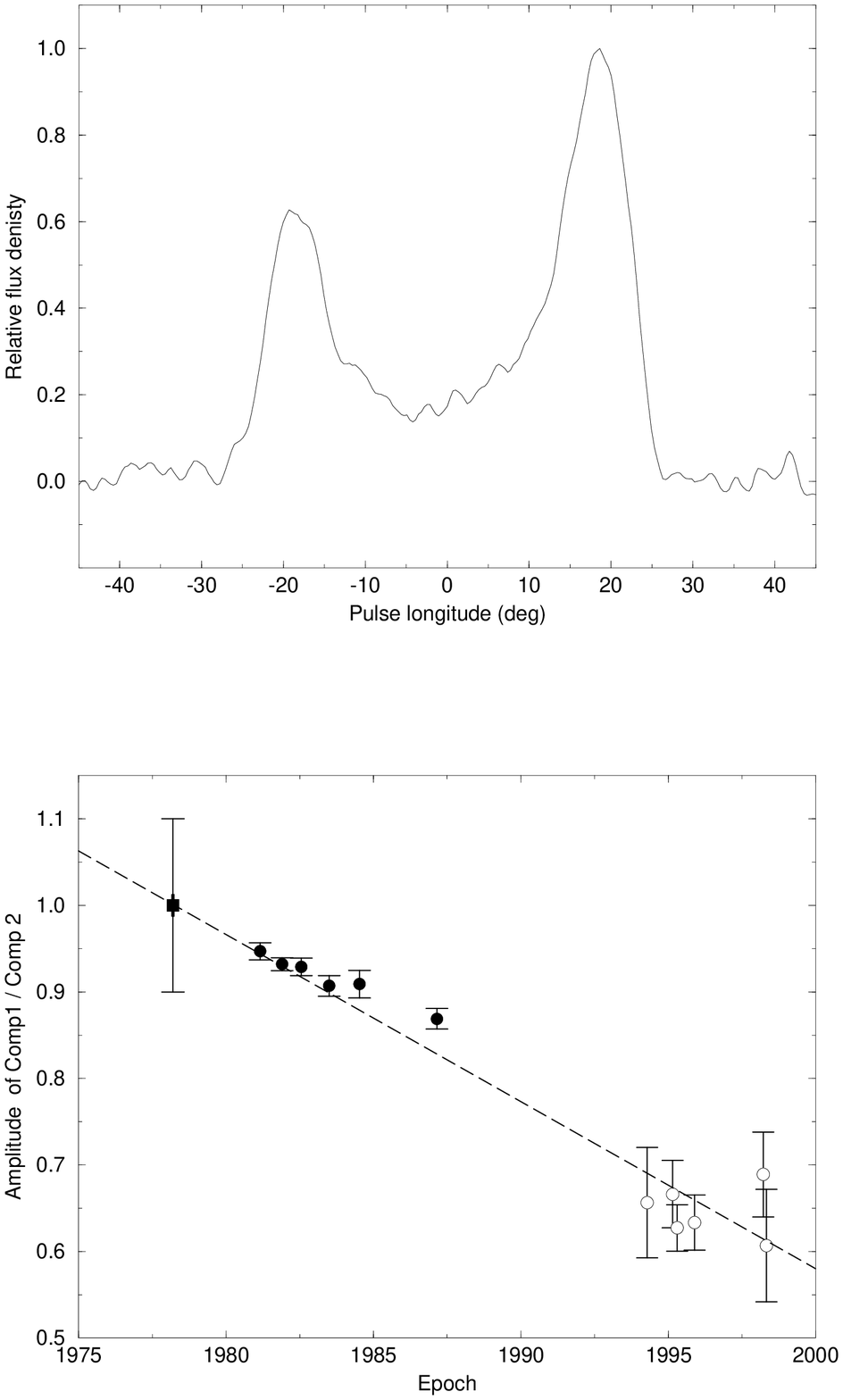} 
\end{tabular}
}

\caption{left) Pulse profile of PSR B1913+16 as presented
by Taylor \& Weisberg (1989). The change in component amplitudes 
studied by Weisberg \& Taylor (1989) is clearly visible, right)
profile of B1913+16 in 1995 and component amplitude ratio as
presented by Kramer (1998).
\label{fig:profiles}}
\end{figure}

Since the precession rate predicted by general relativity is reasonably large,
the pulse profile was naturally studied closely in order to detect possible
changes. Finally, Weisberg et al.~(1989, hereafter 
WRT89)\cite{wrt89}, reported
a change in the relative amplitude of the two prominent 
profile components (Fig.~\ref{fig:profiles}, left).
Analyzing observations made at 1408 MHz from 1981 to 1987, they determined a
change in the amplitude ratio of $1.2\pm0.2$\% yr$^{-1}$, i.e.~a weakening of
the leading component.

These first signs of profile changes could be indeed explained by a changing
cut through the emission beam and hence as the first evidence for geodetic
precession. However, if the emission beam exhibits an overall hollow-cone like
shape, one would also expect a change in the {\em separation} of the two
components rather than only a change in relative intensity. This was not
observed. WRT89 determined an upper limit of $\Delta W<0.^\circ 06$
for a change in component separation in the given time interval. In order to
reconcile this constant profile width with the seen amplitude changes, they
argued that the beam structure is irregular and patchy rather than cone-like,
i.e.~similar to models by Lyne \& Manchester (1988)\cite{lm88}.

With different cuts
through the emission beam, the distance of our
line-of-sight to the magnetic axis should also change with
time. A change in the PA
swing would be expected.  Cordes, Wasserman \& Blaskiewicz (1990,
hereafter CWB90)\cite{cwb90} 
therefore studied polarisation data of PSR B1913+16
to compare profiles and PA swings obtained at frequencies between 1397
and 1416 MHz from 1985 to 1988.  CWB90 neither detected very clear
changes in the pulse shape\footnote{Presumably, this result was caused
by using profiles measured at different frequencies, allowing
frequency evolution to confuse the measurements.}, nor could they find
any significant change in the PA swing. CWB90 pointed out that their
results and those of WRT89 could be still consistent with a hollow
cone emission beam.  They proposed that the existence of a core
component, which is prominent at lower frequencies\cite{tw82} (see
Fig.~\ref{fig:sim}), is causing the change in the component
amplitude ratio.  Finally, they suggested that the lack of detected
changes in profile width and PA swing was due to a special precession
phase at the time of observation.

The situation remained somewhat unsatisfactory. While the amplitude changes
detected by WRT89 seemed to be clear indications of the presence of geodetic
precession, the actually expected behaviour, i.e.~a change in profile width
and PA swing, was not observed.

\subsection{The Times, They are a Changing}

\begin{figure}[t]


\centerline{
\begin{tabular}{c@{\hspace{0.5cm}}c}
\epsfxsize=14pc 
\epsfbox{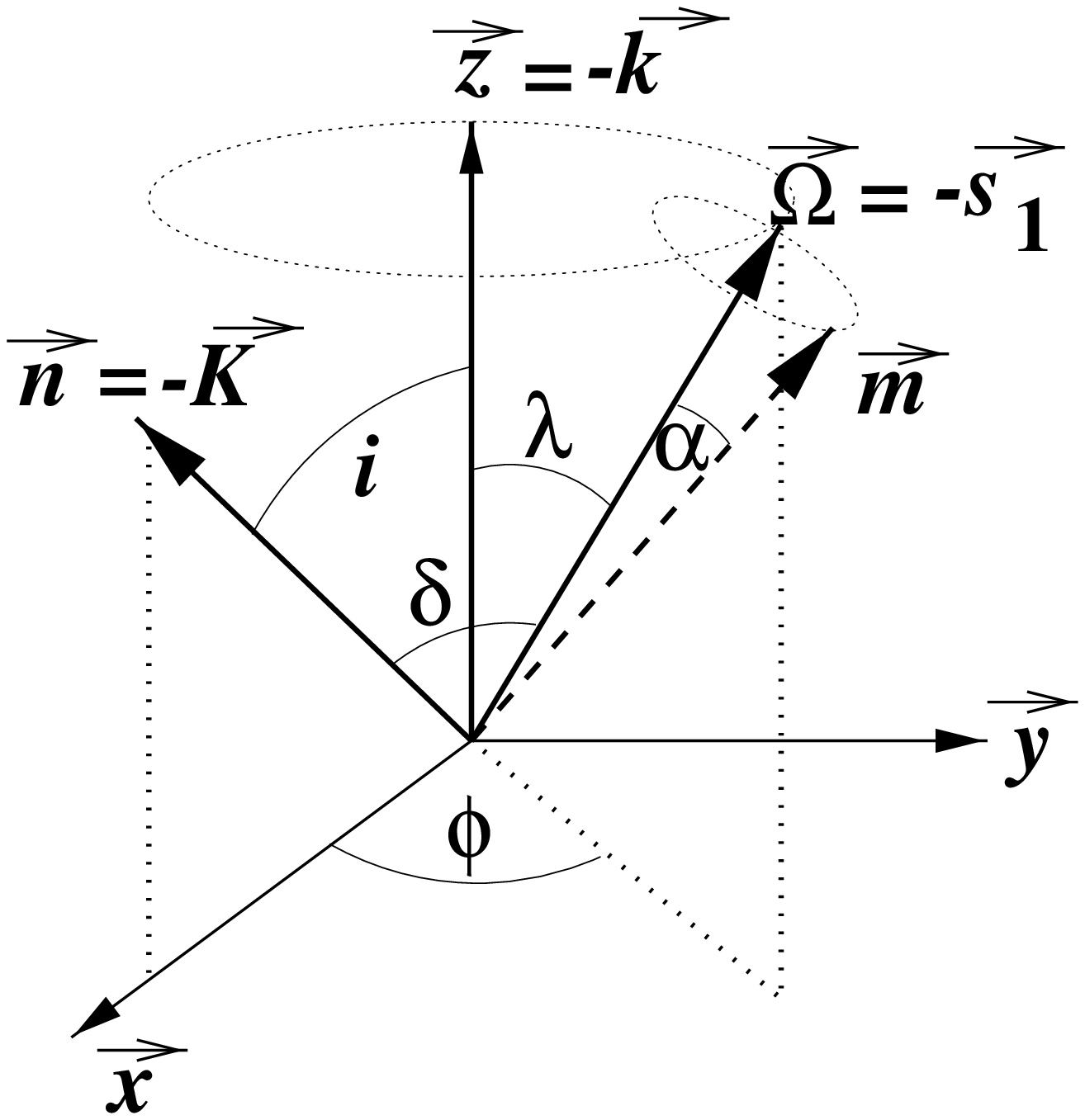} 
\epsfxsize=13pc 
\epsfbox{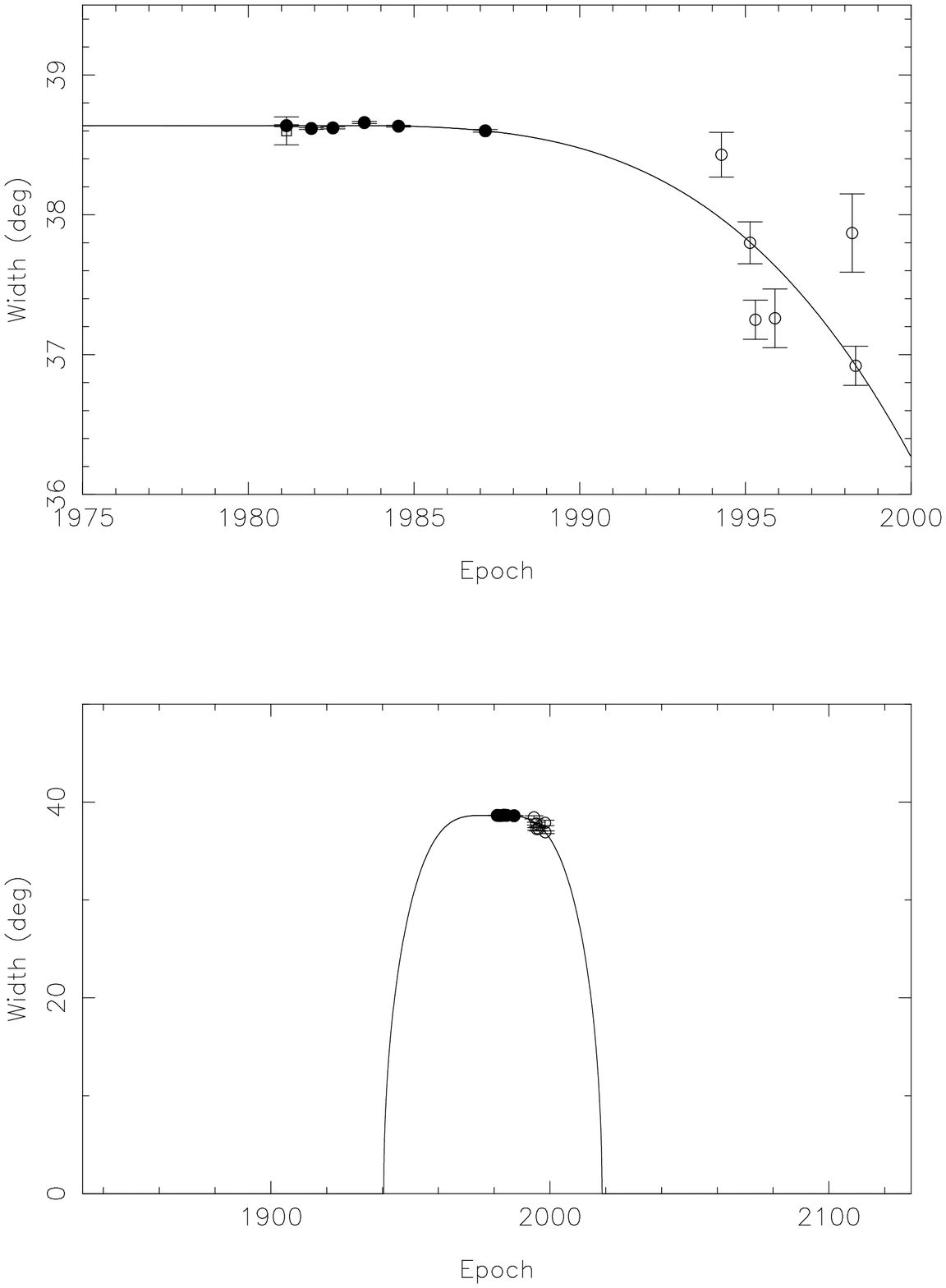} 
\end{tabular}
}

\caption{left) Coordinate system used to describe the performed
modeling by Kramer (1998); right) measurements of the component
separation as a function of time including the best fit (taken
from Kramer 1998).
\label{fig:separation}}
\end{figure}

In 1994 the pulsar B1913+16 was included in a timing program at the 100-m
radio telescope at Effelsberg, which was initiated to take over the regular
timing of millisecond pulsars during the upgrade of the 300-m
Arecibo telescope in
Puerto Rico\cite{wdk+00}. The analysis of profiles measured between 1994
and 1998 by Kramer (1998)\cite{kra98} revealed that the profile components
were still changing their relative amplitude, consistent with the rate first
determined by WRT89 (see Fig.~\ref{fig:profiles}). Even more interesting,
however, was the final detection of a small but significant change in the
separation of the components (see Fig.~\ref{fig:separation}). 
In order to model the long-expected decreasing width of the
profile, two simple assumptions were made, i.e.~those
of a circular hollow cone-like
beam and a precession rate as predicted by general relativity. Both
well justified assumptions (see above) 
lead to a model which has only four free parameters: the misalignment angle
$\lambda$ between the pulsar spin and the orbital angular momentum, the
inclination angle between the pulsar spin axis and its magnetic axis,
$\alpha$, the radius of the emission beam, $\rho$, and the precession phase
given by the reference epoch $T_0$. With these parameters the component
separation at an epoch $t$ is given by\cite{ggr84}
\begin{equation} 
W(t) = 4\sin^{-1} \left[ \sqrt{\frac{\sin^2 (\rho/2) -
\sin^2((\delta(t)-\alpha)/2)}{
       \sin\alpha \sin \delta(t)}} \right] .
\end{equation}
The angle $\delta$ measures the distance of our line-of-sight to the pulsar
spin axis at the closest approach to the magnetic pole. Due to the precession,
this distance will change with time:
\begin{equation} 
\cos \delta(t) = \cos \lambda \, \cos i + \sin \lambda \, \sin i \, 
\cos \phi(t)
\end{equation} 
where $\phi(t)$ is the precession phase,
\begin{equation} 
\phi(t)=\Omega_p\cdot(T_0-t) 
\end{equation} 
With the PK parameters measured by pulsar timing, general relativity allows 
one to
compute the value of $\sin i$, i.e.~the sine of the orbital inclination angle.
In a fortuitous edge-on geometry of the pulsar orbit, we would even be able to
measure this value as the {\em shape} of a Shapiro delay visible in the timing
data (cf.~PSR B1534+12 below). For PSR B1913+16, we simply compute
a value of $i=47.^\circ 2$, or equivalently  $i=180-47.2=132.^\circ 8$ 
(since we only obtain $\sin i$ from timing).

The best fit of this model to the data allows four equivalent solutions. One
pair of solutions corresponds to $i=47.^\circ 2$, the other pair to
$i=132.^\circ 8$, respectively. 
The remaining choice is given by the unknown relative
orientation of the pulsar spin and the orbital angular momentum, i.e.~as to
whether the pulsar rotation is pro-grade or retro-grade. As we will see later,
it can be argued that a retro-grade case is less likely. This leads to the
following best solution\cite{kra98}
\[
i=132.^\circ 8, \; \alpha={153^{+3}_{-8}}^\circ,
\; \lambda={22^{+3}_{-8}}^\circ, \; \rho={9.0^{+4}_{-3}}^\circ
\; \mbox{\rm and} \; T_0=2128\pm4
\]
as this is the only one which gives the right sense of PA swing when compared
with polarization data.\cite{cwb90,bcw91} In fact, this system geometry, which
is solely determined from the change in component separation, is fully
consistent with the results of a RVM fit\cite{bcw91}. Moreover, the obtained
misalignment angle of $\lambda={22^{+3}_{-8}}^\circ$ is in excellent agreement
with earlier simulations\cite{bai88} made to study the number of observable
DNSs which predicted $\lambda\approx20^\circ$ as a typical value for PSR
B1913+16-like systems.

The obtained best fit shown in Fig.~\ref{fig:separation} leads to the
prediction that the pulsar will disappear from the sky around the year
2025! Moreover, it also implies that the component separation remains
almost unchanged for about 60 yr, which corresponds to a likelihood of
20\% to observe the pulsar in that phase given the precession period
of about 300 yr. Looking back, it is now easy to understand why WRT89
were not lucky to detect changes in the component
separation. Similarly, computing the change in PA swing which had to
be measured by CWB90 for a positive detection of a geometry change,
produces a value which is only slightly larger than their estimated
detection limit.

About ten years after the first indications of geodetic precession, it is
finally possible to provide solid evidence for its existence by modelling
profile changes. Most exciting, however, may be the prediction of the pulsar's
disappearance in about 25 years.  Shortly before this, the leading component
will disappear if it continues to weaken with the measured rate (cf.~Istomin
1991)\cite{ist91}.  
Reappearing again around the year 2220, PSR B1913+16 will, in total,
only be observable for about a third of the precession period
(cf.~Figs.~\ref{fig:separation} \& \ref{fig:latest}).

\subsection{Further Results and Update}

\begin{figure}[t]


\begin{tabular}{cc}
\epsfxsize=13pc 
\epsfbox{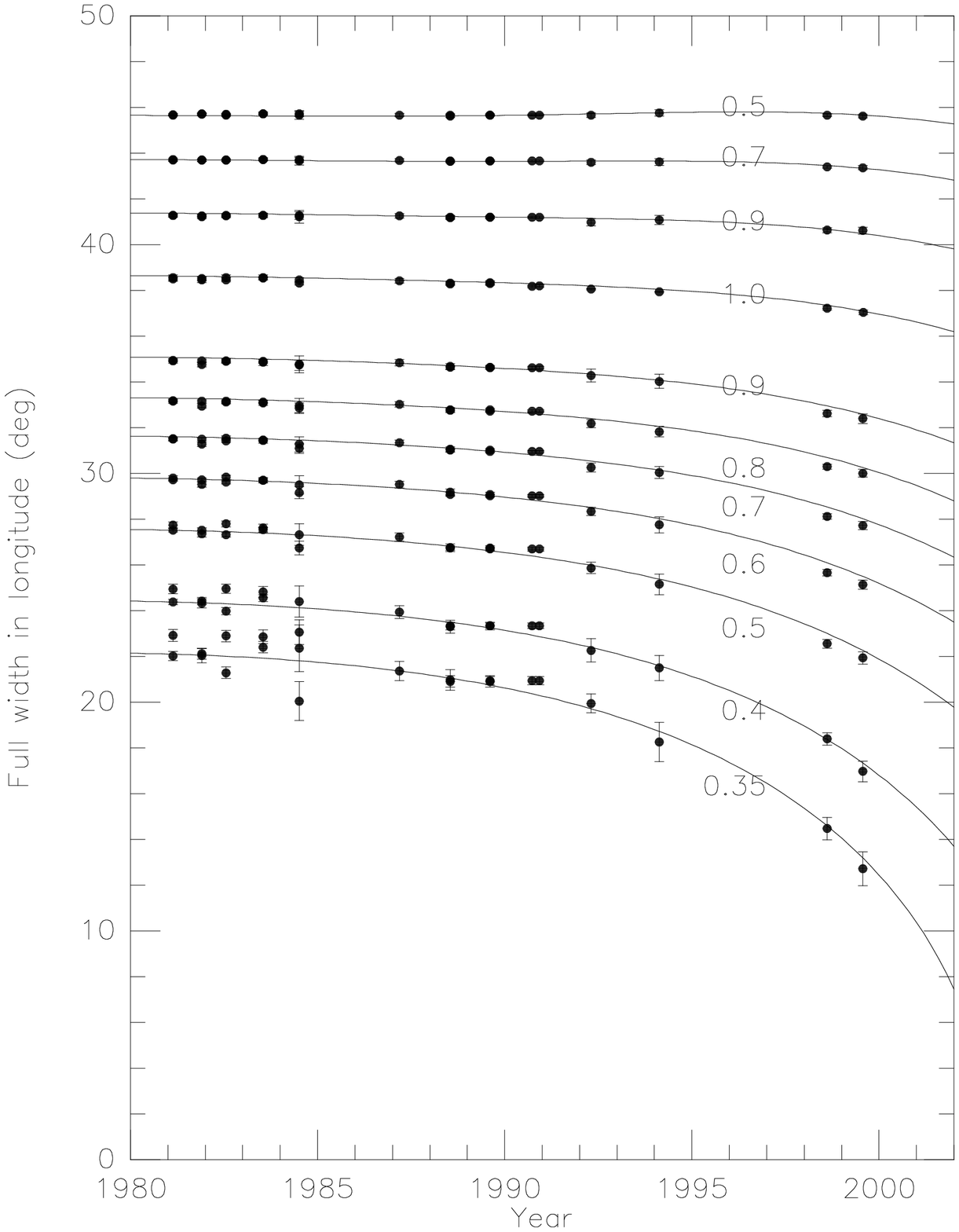} & 
\epsfxsize=16pc 
\epsfbox{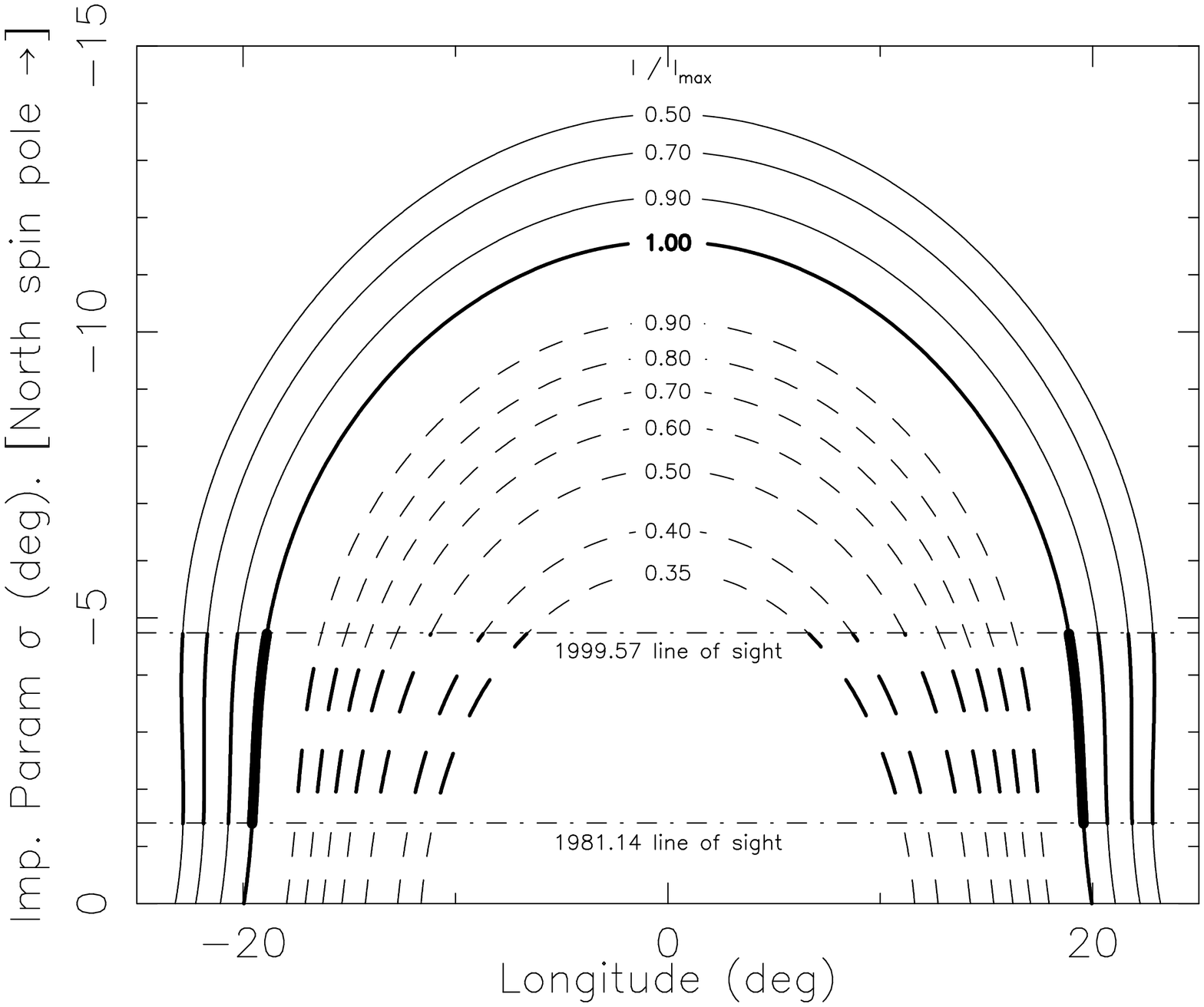} 
\end{tabular}

\caption{left) Decrease of PSR B1913+16's
profile width at different intensity
levels as determined by Weisberg \& Taylor (2000) using the
Arecibo telescope, right)  resulting 2-D pulsar beam map
reconstructed from the different profiles.
\label{fig:weisberg}}
\end{figure}

After the completion of the Arecibo upgrade, Weisberg \& Taylor
(2000)\cite{wt00} obtained new measurements, confirming the results
reported above. Thanks to
the superior sensitivity of the Arecibo telescope, 
they not only measured the same decrease in
component separation, but could also measure a general decrease in
profile width at several intensity levels (Fig.~\ref{fig:weisberg}).
The derived misalignment angle of $\lambda=14(\pm 2)^\circ$ with an
upper bound of $\lambda=22^\circ$\cite{tay99} is in good agreement
with the fit to the Effelsberg data. The obtained data quality even
allows one to do the 
following exciting experiment: Since our line-of-sight
moves through the emission beam, each profiles represents a slightly
different cut through the beam structure.  By adding measured profiles
in the right order, it is possible to reconstruct a real 2-D map of a
pulsar beam for the first time (Fig.~\ref{fig:weisberg}).  The first
results of this mapping procress are surprising: Given the profiles
available so far, the pulsar beam seems to be elongated.
Although the possibility of a non-spherical pulsar beam
has been discussed several times in the literature, even with
claims for elongation in both latitudinal and longitudinal directions,
convincing evidence had not been presented. 

During their data reduction, Weisberg \& Taylor (2000) perform a
brilliant step to separate the measured profiles into odd and even
parts.  While the even profiles are symmetric to a chosen midpoint,
the odd profiles contain all the asymmetries. The addition of both odd
and even parts then reproduces the actual pulse shape. This trick
removes the complicating effects of the change in relative component
amplitude from the symmetric profiles, so that these can be used to map
the 2-D beam structure by applying a specific mapping function.

\begin{figure}[t]


\centerline{
\begin{tabular}{c@{\hspace{-0.5cm}}c}
\epsfxsize=15.5pc 
\epsfbox{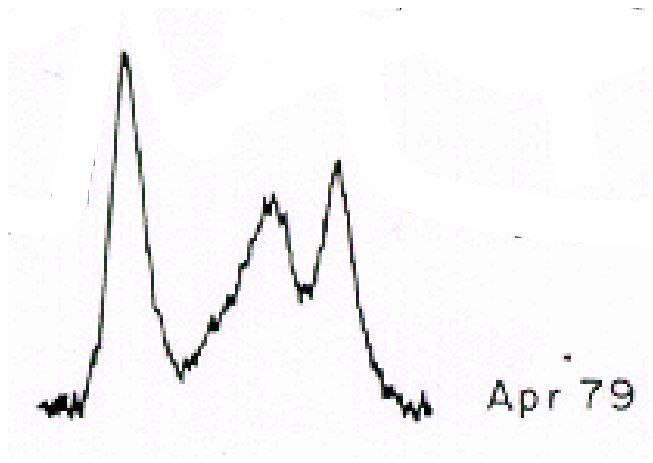} &
\epsfxsize=18pc 
\epsfbox{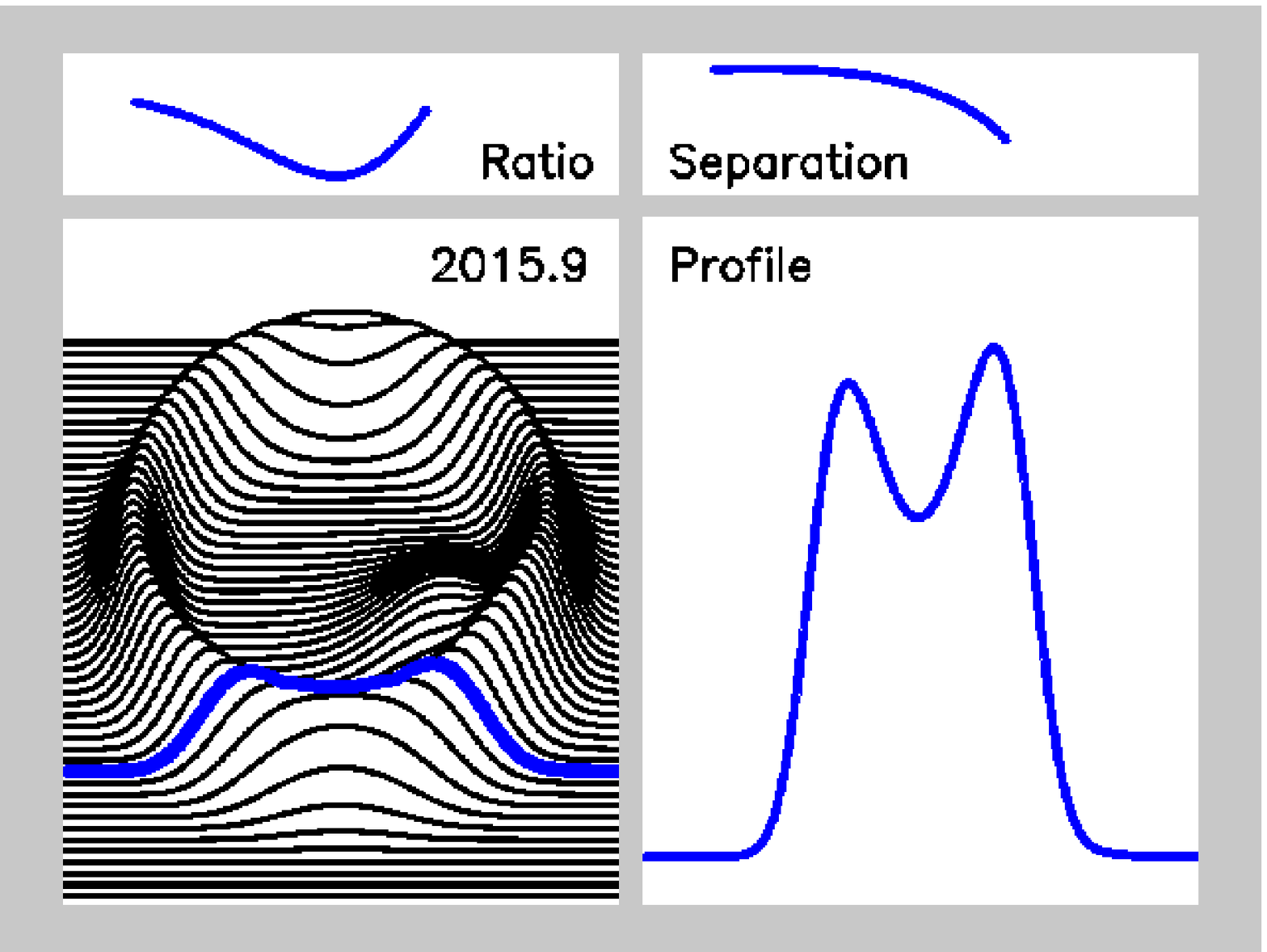} 
\end{tabular}
}

\caption{Further evolution of PSR B1913+16's profile in a simple
hollow cone model including an off-centre core component which is
visible in the 430-MHz profile (left, Taylor \& Weisberg 1982).  
The right
figure is a snapshot of a simulation measuring component amplitude
ratio and separation.  While the component amplitude ratio first
decreases as observed, we can predict that it will rise again as soon
as the line-of-sight leaves the core dominated region.
\label{fig:sim}}
\end{figure}

Since an elongation of the pulsar beam would have interesting
implications (for instance for birth rate calculations), it is
stimulating to compare the results of Weisberg \& Taylor (2000) with
the outcome of an independent study that employs reverse
engineering (Kramer 2001, in prep.).  Starting with a {\em
circular} cone, cuts through the emission beam are computed for 
different lines of sight.  Using the resulting profiles, a beam map is
computed, following the same procedure as applied by Weisberg \&
Taylor (2000). We are interested to see whether we are able to
reproduce the observed profiles without resorting to elongated beams.
Detailed results of this study will be presented elsewhere, but one
interesting effect can already be obtained: in order to produce the
asymmetric profiles as observed in reality, a core component is placed
in the beam at a slightly off-centre position. The core is visible at
low frequencies (see Fig.~\ref{fig:sim}), and its possible influence was
already pointed out earlier\cite{t79,cwb90}.  Using the
profiles simulated for given epochs, we can determine both the
expected component amplitude ratio and the component separation and
compare them with the observations. A snapshot of these simulations is
shown in Fig.~\ref{fig:sim}. As a result, we can not only reproduce
the observed values, but we can even make a prediction: when the
line-of-sight leaves the central region dominated by the core, the
amplitude ratio should increase again and may eventually reach unity
again (Kramer 2001, in prep.).

\begin{figure}[t]


\centerline{
\begin{tabular}{cc}
\epsfxsize=11pc 
\epsfbox{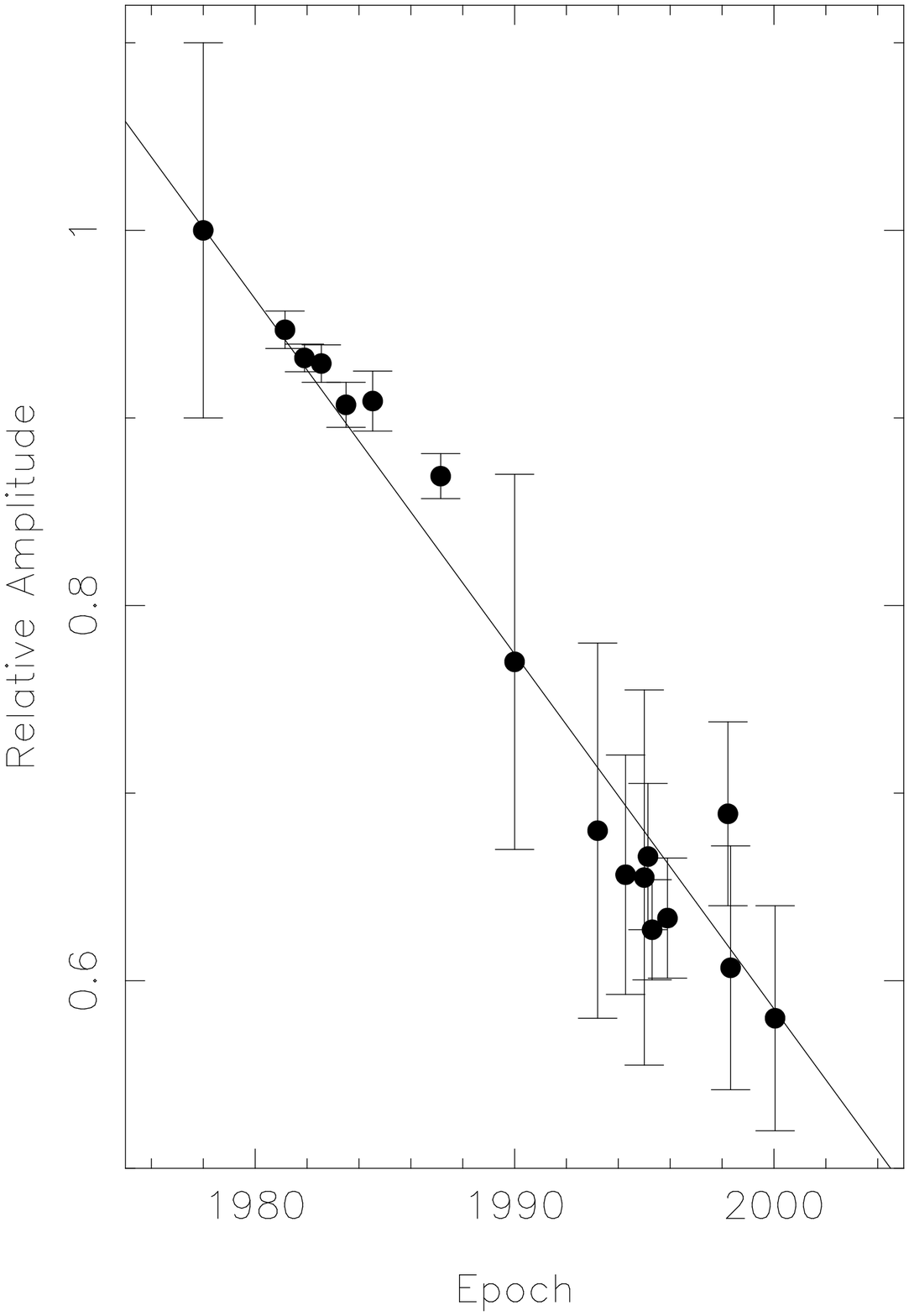} & 
\epsfxsize=11pc 
\epsfbox{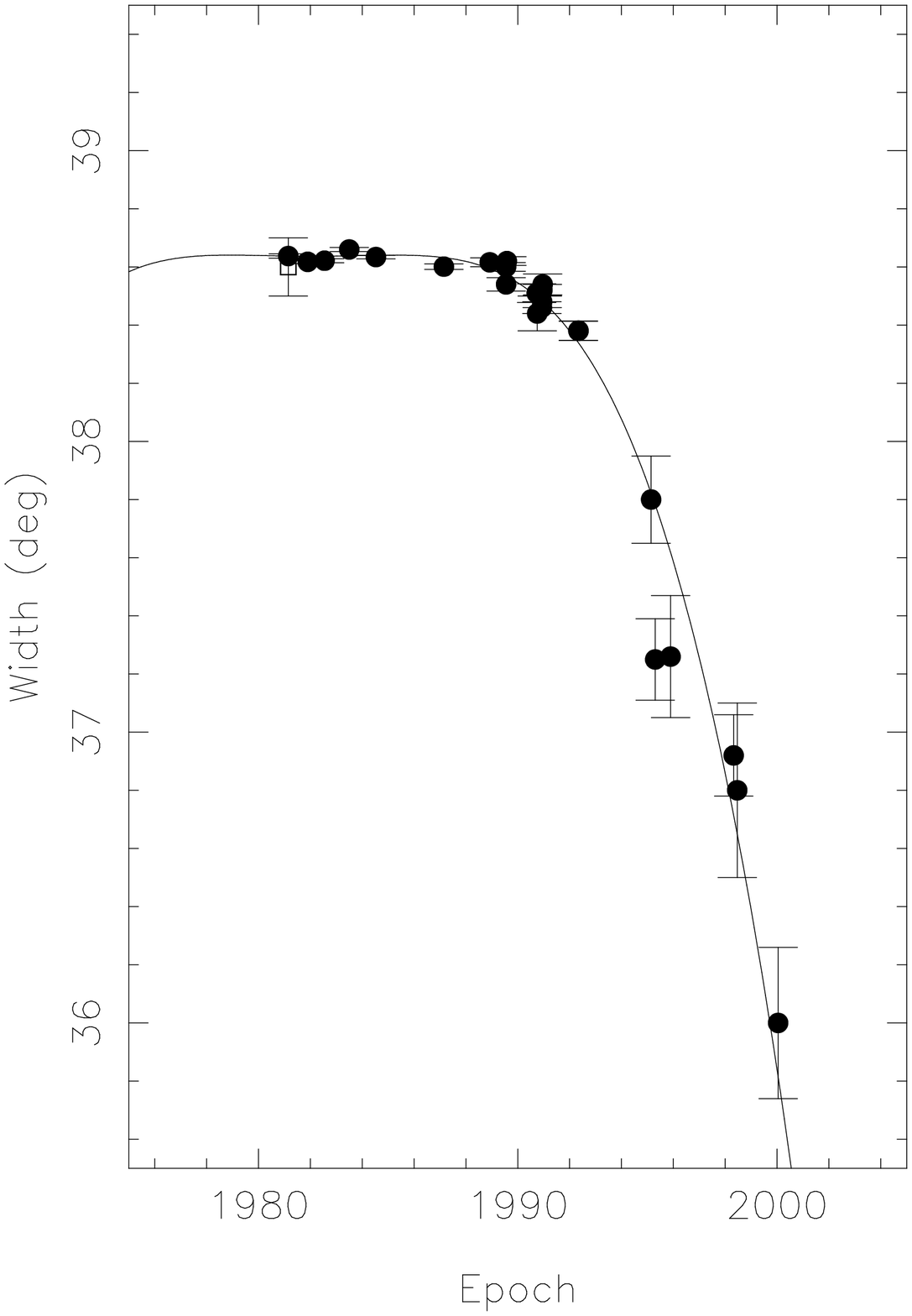} 
\end{tabular}
}

\caption{Latest results for measurements of the still
decreasing component amplitude ratio (left) and the
decreasing component separation (right). The left plot
contains unpublished data from Jodrell Bank, while the
right plots includes previously unpublished data
by Weisberg \& Taylor (2000). During the fit leading to
the results shown on the right, we treated $\Omega_p$ as a
free parameter.
\label{fig:latest}}
\end{figure}

In Fig.~\ref{fig:latest} we present the latest measurements obtained
with the Effelsberg telescope. We also add Weisberg \& Taylor's
previously unpublished Arecibo data. As visible from the new
measurements, the previously observed trend continues. In the coming
years, the decrease in profile width should accelerate drastically,
eventually allowing us to turn this qualitative test of general
relativity into a quantitative one. For demonstration, we attempted to
treat the rate of precession as a free parameter rather than keeping
it fixed to the value predicted by general relativity.  The result,
$\Omega_p=1.2\pm0.2$ deg yr$^{-1}$, (shown in Fig.~\ref{fig:latest})
is by no means a precision test and not even a proof of principle! In
order to succeed, one has to be confident that the beam structure is
modelled correctly. Certainly, this will be the most difficult 
part of the job. However, it may be possible
that an iterative process will be finally successful.  Moreover, we
did not make much use of the available polarisation information
yet. While we used PA data to exclude certain classes of models, we
did not invoke RVM fits because of the mentioned large
uncertainties. Karastergiou et al.~(2000)\cite{aris00} presented a method
where polarisation and pulse shape data can be combined without
relying on RVM fits which will become very useful. 

Combining the polarisation data will be a crucial test, as we
should definitely detect changes in the PA swing finally.  Eventually,
one may obtain reliable measurements of $\Omega_p$.  However, the most
reliable test would be, of course, the re-appearance of the pulsar at
about 240 years after it disappears --- if we can wait that long.

\section{Asymmetric Supernova Explosions}

Geodetic precession occurs if the pulsar spin is misaligned with the
orbital momentum vector. For PSR B1913+16, we can measure this
misalignment angle using the observed precession effects to obtain
$\lambda\approx22^\circ$. The question about the origin of this angle
is far from trivial. In fact, a non-zero $\lambda$ is the
imprint of the violent birth event which created the DNS.  It is the
ultimate proof that supernova explosions are asymmetric.

\subsection{The Birth of a DNS}

\begin{figure}[t]


\centerline{
\begin{tabular}{cc}
\epsfxsize=15.5pc 
\epsfbox{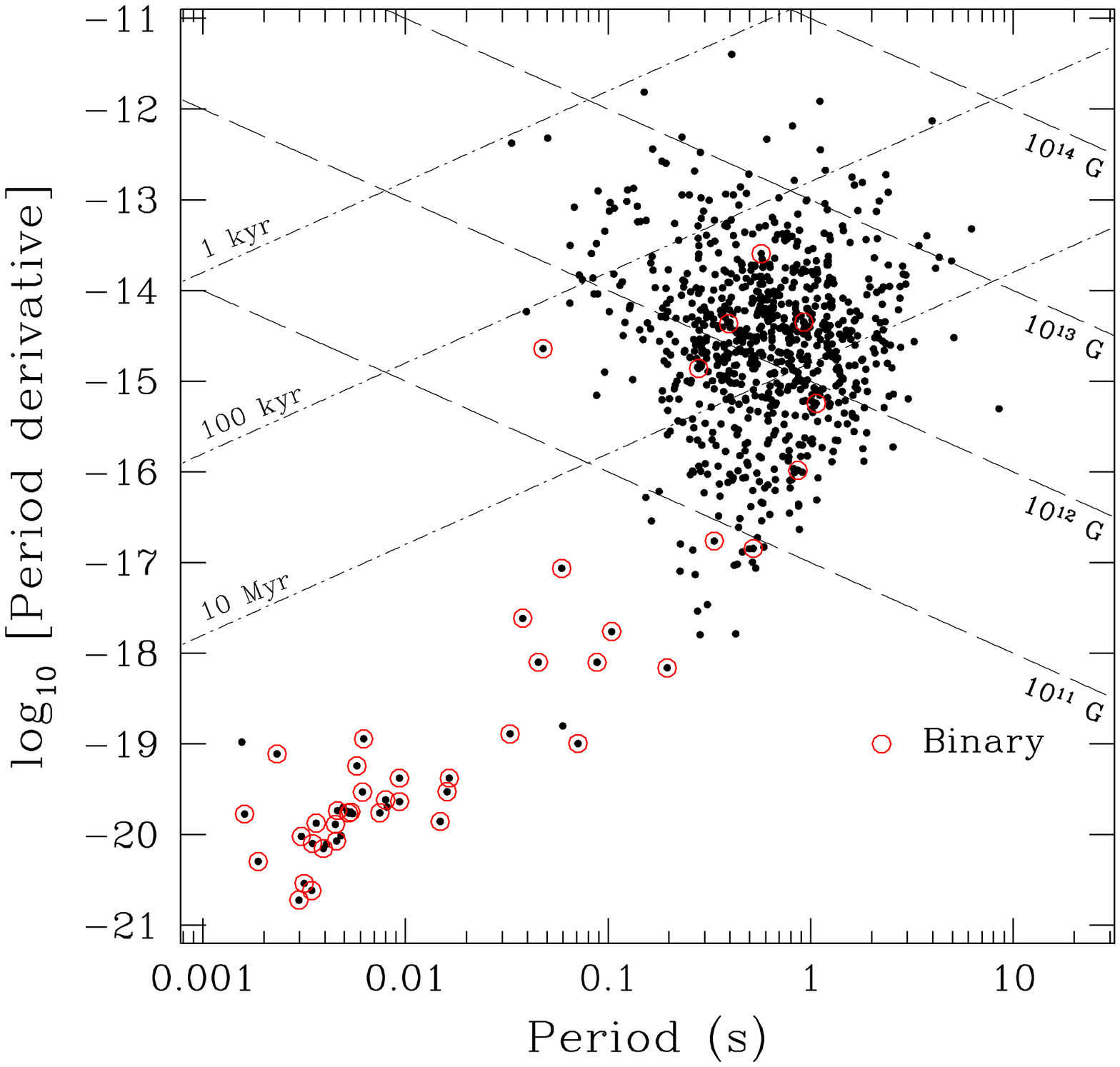} & 
\epsfxsize=6pc 
\epsfbox{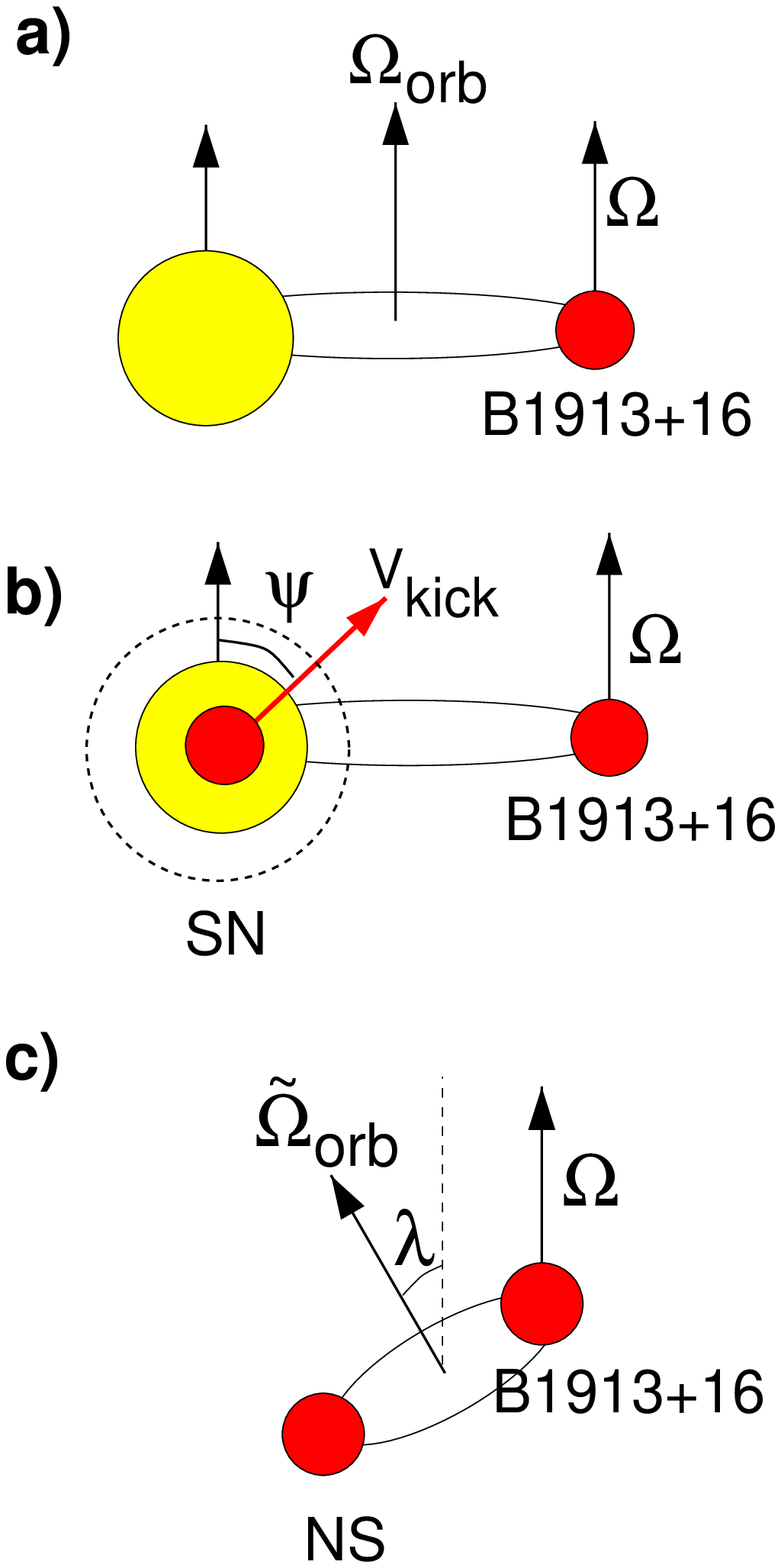} 
\end{tabular}
}

\caption{left) $P-\dot{P}$-diagram as known at the time of writing.
Binary pulsars are marked with circles. Lines of constant age and
surface magnetic field strengths are indicated; right) evolutionary
stages of the PSR B1913+16 (a) before and (c) after the second
supernova explosion.  Before the explosion all spin vectors are
aligned. After the explosion, imparting a kick to the companion, the
new orbit is misaligned to the previous orbit.
\label{fig:ppdot}}
\end{figure}

In the beginning, a DNS starts off as a massive binary system where
the more massive companion eventually explodes and becomes a
pulsar. The pulsar is probably born with an initial period of 10--30
ms and a high spin-down rate $\dot{P}$ (upper left region
in $P-\dot{P}$-diagram,
Fig.~\ref{fig:ppdot}).  As the rotation slows down to a period of a few
seconds, the pulsar emission eventually ceases after tens of 
million years or so, moving the pulsar to the lower right corner
of the  $P-\dot{P}$-diagram. 
When the massive companion evolves and overflows its Roche-Lobe,
the binary system reaches a High-Mass X-Ray Binary phase. The dead
pulsar accretes matter and angular momentum. This spin-up ``recycles''
the pulsar, which re-appears as a millisecond pulsar in the lower left
part of the $P-\dot{P}$-diagram. The spin period achieved
during the recycling process depends on the
duration of the mass transfer. For low-mass systems this phase can be
rather long, resulting in periods of a few ms as usually observed for
pulsars with white-dwarf companions. In the case of high-mass systems
however, this phase is ultimately limited by the event of the
supernova explosion (SN) of the companion.

Just prior to the explosion of the companion, the binary system is
expected to consist of a recycled pulsar in a circular orbit about its
companion, which has evolved to a He-star\cite{fk97}. All spin vectors are
aligned as a result of the angular moment transfer during the accretion
process and a following common envelope phase. When the He-star
explodes, the pulsar itself is not directly affected by the explosion,
since its cross section is too small. In contrast, the survival of the
binary system depends on a number of lucky circumstances.

If the explosion is symmetric, the survival will be mainly determined
by the total mass loss. Reality, however, is more complicated, since
there is convincing evidence that SNe are in fact asymmetric. In that
case, a momentum ``kick'' is imparted to the newly born neutron
star. Unless the magnitude and direction of the kick are favorable, 
the binary system disrupts\cite{fk97}.


\subsection{The Kick --- Lessons from Geodetic Precession}

The actual process causing the kick is still poorly understood. Most of the
proposed models, which can be roughly classified as electromagnetic kicks,
neutrino driven kicks or hydrodynamical kicks, either produce velocities too
small compared to observations, or require unrealistic conditions such as
extremely large magnetic fields\cite{lai99}. It turns out that it is rather
difficult to produce kick velocities in excess of, say, 300 km s$^{-1}$.  In a
  recent work, Wex et al.~(2000) have used the system information obtained by
  geodetic precession for PSR B1913+16 to derive constraints on the kick
  mechanism\cite{wkk00}.

Since the recycled pulsar is unaffected by the explosion of its companion,
its spin vector will continue to point in the direction of the pre-SN
orbital angular momentum. The new orbital plane, and hence the new orbital
angular momentum, is determined by the direction and magnitude of the kick and
will be usually different from the old orbit. As a result, the two vectors are
misaligned and gravitationally coupled -- geodetic precession occurs
(Fig.~\ref{fig:ppdot}).

\begin{figure}[t]


\centerline{
\epsfxsize=21pc 
\epsfbox{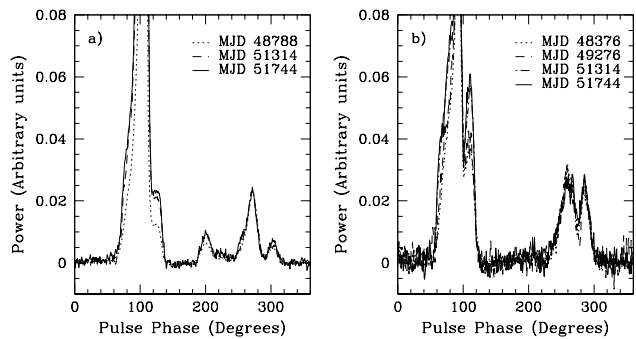} 
}

\caption{Profiles of PSR B1534+12 as measured by Stairs et al.~(2000)
at 430 MHz (left) and 1400 MHz (right).  Changes due to geodetic
precession are clearly visible.
\label{fig:1534}}
\end{figure}

It is possible to write down a set of equations relating the pre- and
post-SN orbit using conservation of momenta, conditions for the orbit,
typical sizes of He-star etc\cite{fk97}. Knowing that all spin
vectors before the SN had been aligned, and assuming that the system
velocity after the second SN is dominated by the effects of the kick,
we can infer those kick parameters that are necessary to explain the
post-SN system characteristics. This procedure is possible for PSR
B1913+16 (Wex et al.~2000)\cite{wkk00}: 
Because of its nature as a DNS, both stars do
not interact, but the system's evolution is solely determined by the
emission of gravitational waves and its motion in the galactic
potential. Both can be calculated backwards in time. The (justified)
assumption that the system was born in the galactic plane, marks the
termination point of this backwards computation. The derived orbit and
system velocity must then be similar to the conditions just after the
SN.

With the unmeasurable radial velocity as the only free parameter, only
three different cases are consistent with the age of the pulsar and
the system's current motion away from the plane. Considering all these
cases and possible pro- and retro-grade solutions, Wex et al.~(2000)
reach a number of interesting conclusions among which are: a) on the
average, a larger kick is required to explain a retro-grade solution;
b) kick velocities between 300 and 1100 km s$^{-1}$ are necessary; c)
independent of radial velocity and considered cases, the kick imparted
to the companion must have been almost perpendicular to the orbital
spin. Given the alignment of all pre-SN spins, the kick was also
perpendicular to the rotation axis of the exploding star. That has
very interesting implications. Firstly, the duration of the kick must
have been short compared to the rotational period of the exploding
star (otherwise this velocity component would have
disappeared). Secondly, if we assume that the magnetic field of the
companion was more or less aligned with the rotational axis, as it is
the case for the Earth or Sun, the kick was also perpendicular to the
magnetic axis. The latter rules out all kick mechanisms which have a
preference in the direction of the magnetic field, as for instance
some neutrino oscillation models\cite{lai99}.

Yet again, pulsars can obviously serve as diagnostic tools for studying a
rather different branch of astrophysics.  By carefully investigating this
particular binary system, we are able to derive important constraints for any
kick mechanism proposed, i.e.~every proposed model should be able to produce a
short kick perpendicular to the rotation axis.

\section{Evidence for Geodetic Precession --- Part II.}

\subsection{Geodetic Precession and PSR B1534+12}

The next obvious source to look for evidence of geodetic precession is
the second DNS discovered in the disk of our Galaxy, 
PSR B1534+12\cite{wol91a}. This 38-ms pulsar
happens to be brighter than PSR B1913+16, exhibiting also a much
narrower pulse peak which both lead to a much better timing
accuracy. The fortunate edge-on orientation of the orbit allows the
measurement of two PK parameters describing the Shapiro delay of the
pulse arrival times. In total five PK parameters can be
measured\cite{sac+98} (see Stairs, these proceedings).
With
the mass estimates of $m_p=1.334\pm0.002${} and $m_c=1.344\pm0.002$
(Table \ref{tab:parms}), the
expected precession rate of $\Omega_p=0.51^\circ$ yr$^{-1}$ is much
lower than that of PSR B1913+16 (Fig.~\ref{fig:rates}). Nevertheless,
Arzoumanian (1995)\cite{arz95} and Stairs et al.~(2000)\cite{stta00} 
succeeded in detecting 
secular changes in the pulse profiles providing further evidence for
geodetic precession acting in binary pulsars (Fig.~\ref{fig:1534}).

The profiles of recycled pulsars with periods less then 50 ms do not
seem to fit the classical hollow-cone beam model anymore\cite{kll+99}.
This may stem from the extreme compactness of millisecond pulsar
magnetospheres, whose sizes scale with the pulse period. The
difficulties in modelling the profile changes observed for PSR
B1534+12 in a manner done for PSR B1913+13 are hence unfortunate but
not too surprising. The situation can possibly be saved by the
excellent polarisation data available. Since emission can be seen for
almost 80\% of the pulse period, RVM fits provide information about
the viewing geometry of PSR B1534+12 with an accuracy much better than
for most other pulsars\cite{stta00}.

\subsection{PSR J1141$-$6545 --- Further Evidence?}

The latest relativistic binary discovered in the Parkes Multibeam
Survey\cite{klm+00a}, PSR J1141$-$6545, was first considered to be a
DNS. The short orbital period of 4.7 hours and an orbital eccentricity
of $e=0.171$ (Tab.~\ref{tab:parms}) fit well into the range known
for DNSs. Soon after its discovery, the first PK parameter could be
measured, i.e.~the periastron advance, $\dot\omega$. Assuming that its
value is completely determined by relativistic effects, general
relativity then allows one to determine the total mass of the
system\cite{dt92}, $M=m_p+m_c=2.4$.  Assuming a canonical pulsar mass
of $m_p\approx1.4${} leaves only $m_c\approx 1$ for the
companion. These estimates lead to the conclusion that the companion
is almost certainly a heavy white dwarf\cite{klm+00a}. 

\begin{figure}[t]


\centerline{
\begin{tabular}{cc}
\epsfxsize=13pc 
\epsfbox{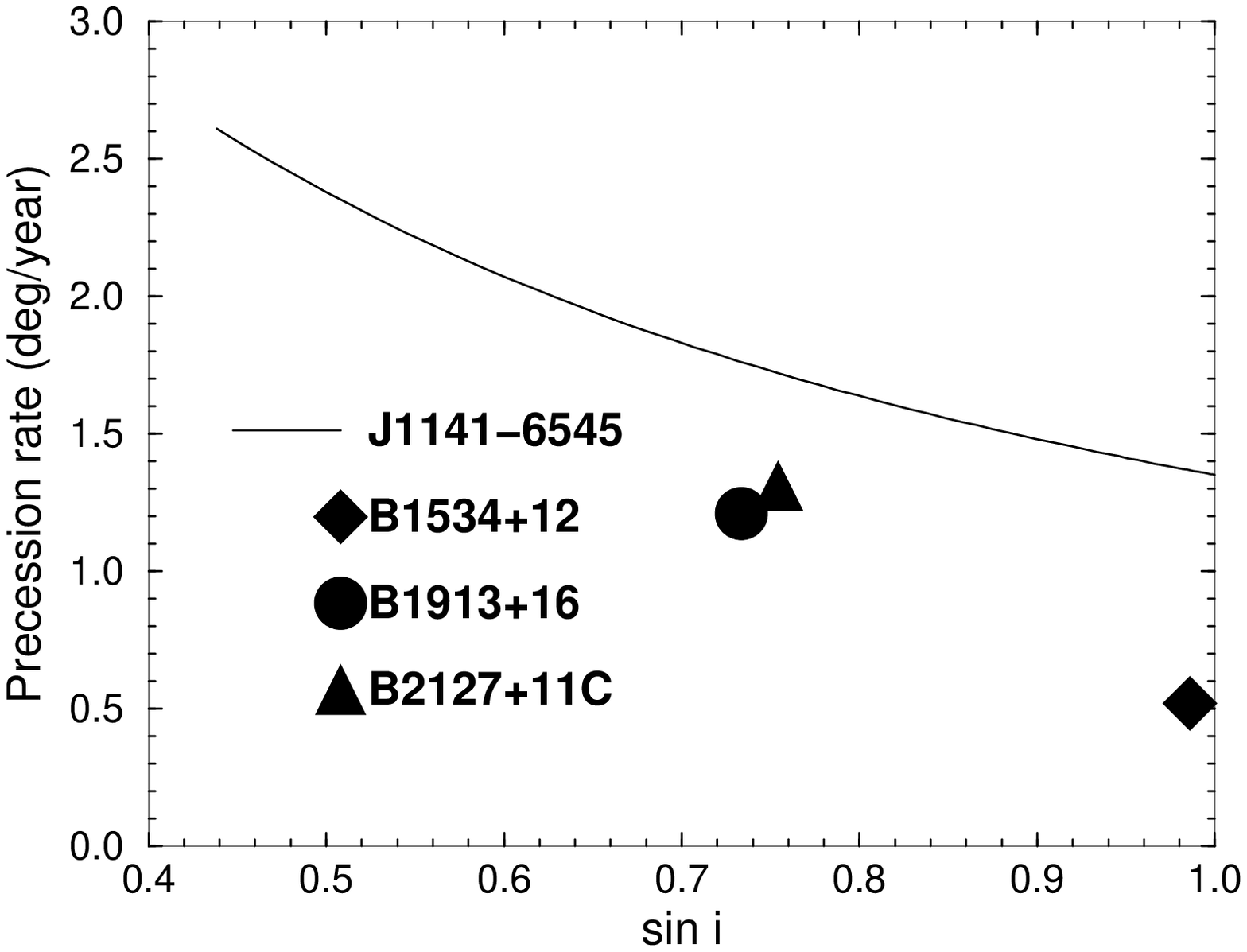} & 
\epsfxsize=11pc 
\epsfbox{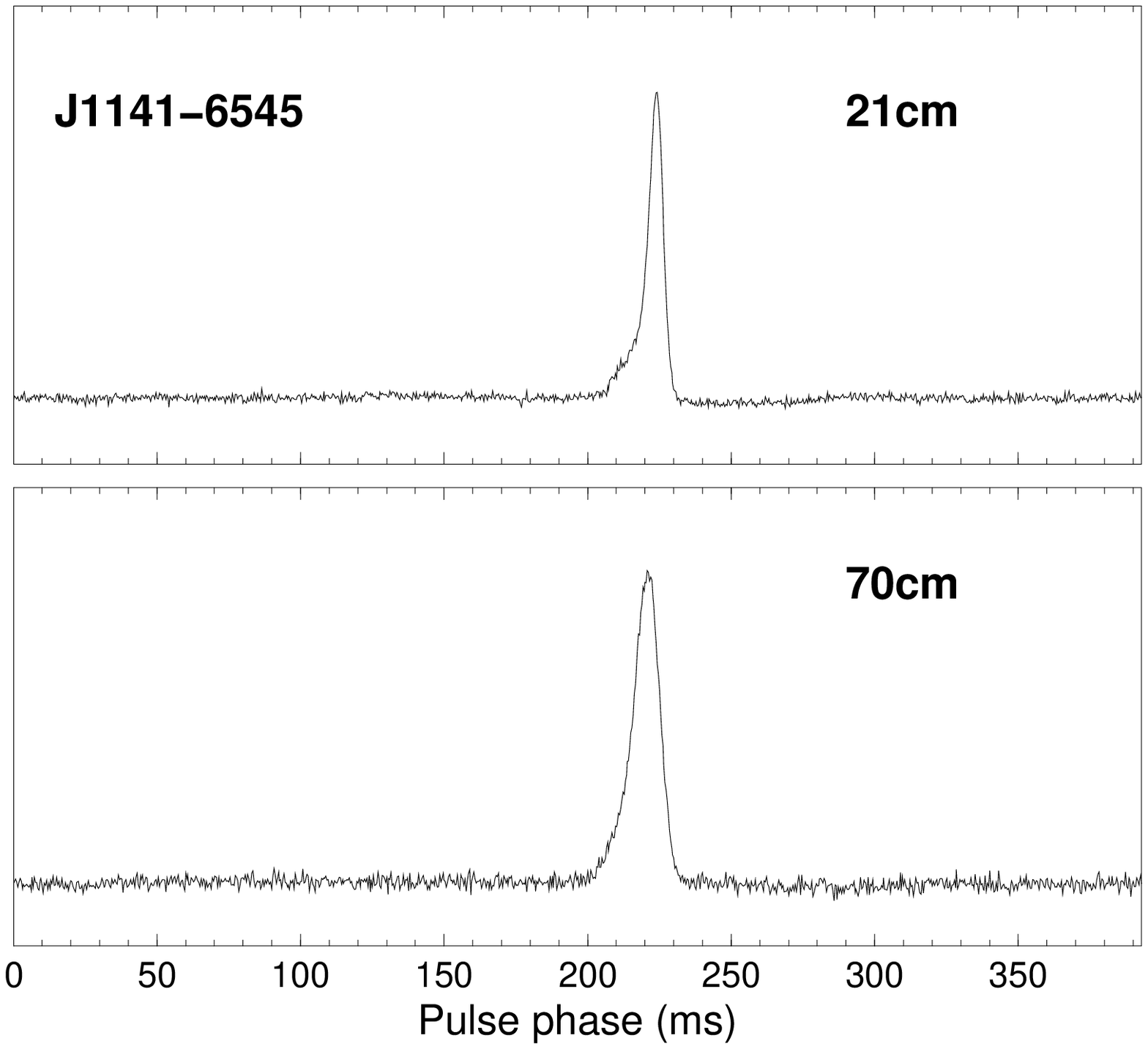} 
\end{tabular}
}

\caption{left) Precession rates as predicted by general relativity
for the sources discussed here and a further DNS PSR B2127+11C. Unfortunately,
its pulsar is rather weak. The precession
rate for PSR J1141$-$6545 is given as a function of the unknown orbital
inclination angle; right) profiles of PSR J1141$-$6545 at 430 and
1400 MHz (Kaspi et al.~2000).
\label{fig:rates} }
\end{figure}

The exact mass distribution in the system can be determined if a
second PK parameter is measured, which should be easily possible in
the future.  For the moment, we can make use of the mass function of
the binary system which indicates that the inclination angle must be
close to $i=90^\circ$ to produce a pulsar mass of $m_p=1.4$. Using
these values, the predicted precession rate is the largest for any
known system, i.e.~$\Omega_p\approx1.35$ deg yr$^{-1}$. For smaller
(though unlikely) inclinations, the value even increases (see
Fig.~\ref{fig:rates}).

Obviously, even as a pulsar-white dwarf system, PSR J1141$-$6545 is
a prime candidate to exhibit geodetic precession. It turns out that
the same spot of sky had been surveyed at least twice ten years
ago, but no pulsar was discovered, neither at 430 nor at 1400 MHz.
Today, the pulsar is fairly strong at both frequencies, with a flux
density several times that of the detection limit of the previous
surveys (Fig.~\ref{fig:rates}).  Unless other effects like man-made
radio interference are responsible for the previous non-detections, it
is conceivable that the pulsar is indeed precessing and has moved its
beam into our line-of-sight during the past ten years. If that is the
case, future observations should show profile changes soon. Since the
pulsar is strong and also polarised, we can hopefully 
present further detailed evidence for geodetic precession in the
near future.

\section{Outlook}

It is clear from the presented observational evidence that general
relativity correctly predicts the existence of geodetic precession.
While the presently available data provide only qualitative tests so
far, quantitative tests may be possible if the pulsar emission beam
can be modelled sufficiently. Alternatively, polarisation data may
produce reliable geometry information via RVM fits, so that a
measurement of $\Omega_p$ will be possible.

Although the analysis of pulse structure data is the most approachable
method to investige geodetic precession, recently Doroshenko et
al.~(in prep.) investigated the effects on pulsar timing in more
detail. It turns out that geodetic precession may produce after all a
measurable effect in the timing. Applying their calculations to PSR
B1913+16 leads however to the conclusion that this effect may be
absorbed in the fits for $P$ and $\dot{P}$, since the pulsar may only
be visible for 60 to 80 years. It has yet to be determined as to 
whether other pulsars could be useful.

As we have seen, one major consequence of geodetic precession can be
the (temporary) disappearance of a known pulsar from the sky.  Over a
time span that is long compared to typical precession periods,
i.e.~several hundreds to thousands of years, more DNSs will be
discovered and measured than without geodetic precession. However, for
the typical life time of a scientist, the relevant numbers will not
change: the number of binary pulsars that will disappear from the sky
before we can detect them should be similar to the number of sources
turning up for the first time. In other words, we live in a
``steady-state precessing pulsar universe'', so that birth rate and
hence detection calculations for gravitational wave detectors are
unaffected by the discussed phenomena. While this consideration
certainly applies in general, it is a very reasonable idea to
occasionally re-search the position of known DNSs with large
precession rates to check for the appearance of previously unseen
companions. You never know...

\section*{Acknowledgments}
It is a pleasure to thank the organisers for a remarkable conference!
I would like to thank Norbert Wex and Vicky Kalogera for the great
collaboration on pulsar kicks. I am grateful to Don Backer, Ingrid
Stairs, Joe Taylor, Joel Weisberg and in particular to Norbert Wex for
very useful stimulating discussions. I also thank Ingrid for
providing me with figures, a careful reading of the manuscript and
valuable exchange of ideas. It is a pleasure to thank Andrew Lyne, Francis
Graham-Smith and in particular Duncan Lorimer for valuable comments
on the manuscript!



\end{document}